\documentclass[pre,eqsecnum,showpacs,amssymb]{revtex4}
\usepackage{amsmath,graphicx}
\usepackage{epsfig}
\usepackage{amssymb}
\usepackage{color}

\begin{document}

\title{Dynamic Compression of {\it in situ} Grown Living Polymer
Brush: Simulation and Experiment}

\author{K. Jalili$^{1,2}$, F. Abbasi$^2$, A. Milchev$^{1,3}$}
% and T.A. Vilgis$^1$}
\affiliation{$^1$ Max Planck Institute for Polymer Research
  10 Ackermannweg, 55128 Mainz, Germany\\
$^2$ Institute of Polymeric Materials, Sahand University of Technology P.O.Box
51335-1996, Tabriz, Iran\\ $^3$ Institute for Physical Chemistry, Bulgarian
Academy of Science, 1113 Sofia, Bulgaria}
\begin{abstract}
A comparative dynamic Monte Carlo simulation study of polydisperse living
polymer brushes, created by surface initiated living polymerization, and
conventional polymer monodisperse brush, comprising linear polymer chains,
grafted to a planar substrate under good solvent conditions, is presented. The
living brush is created by end-monomer (de)polymerization reaction after placing
an array of initiators on a grafting plane in contact with a solution of
initially non-bonded segments (monomers). At equilibrium,  the monomer density
profile $\phi(z)$ of the LPB is found to decline as $\phi(z) \propto
z^{-\alpha}$ with the distance from the grafting plane $z$, while the
distribution of chain lengths in the brush scales as $c(N) \propto N^{-\tau}$.
The measured values $\alpha \approx 0.64$ and $\tau \approx 1.70$ are very close
to those, predicted within the framework of the Diffusion-Limited Aggregation
theory, $\alpha = 2/3$ and $\tau = 7/4$. At varying mean degree of
polymerization (from $\langle L \rangle = 28$ to $\langle L \rangle = 170$) and
effective grafting density (from $\sigma_g = 0.0625$ to $\sigma_g = 1.0$), we
observe a nearly perfect agreement in the force-distance behavior of the
simulated LPB with own experimental data obtained from colloidal probe AFM
analysis on PNIPAAm brush and with data obtained by Plunkett {\it et. al.},
[{\it Langmuir} {\bf 2006}, 22, 4259] from SFA measurements on same
polymer.

\end{abstract}

\maketitle

\section{Introduction}
\label{sec_intro}

After the recent advances in the nanostructuring of surfaces, polymers can now
be end-grafted on nanometer scale structures \cite{Dyer,Gates}. The recent
progress in the production of nanoscale polymer structures has, however, only
partially been matched by advances in the experimental methods of measuring the
properties of such systems \cite{Kawaguchi}. Available contactless methods, for
example, optical methods, average the brush properties over areas that are
larger than the nanoscale so that fine spatial resolution is lost. To retain
sufficient spatial resolution, atomic force microscopy (AFM) is conventionally
used \cite{Binnig,Magonov,Garcla,Kaholek}. However, the necessary contact
between the AFM tip and the polymer thin layer distorts the layer and thus the
object that it intends to study \cite{Wilder}.

AFM measurements of  polymers, tethered on surfaces with various geometries,
illustrate a more general and fundamental class of problems, namely, the
response of soft matter to external forces. Polymer brushes, that is, densely
packed arrays of polymer chains end-attached to an interface, have been studied
extensively (for reviews see
\cite{Halperin1,Klein1,Szleifer,Grest1,Kaholek,Binder}) due to their ability to
modify surface properties, prevent colloid aggregation, and enhance lubrication
or adhesion \cite{Kaholek,Napper,Hamilton,Pastorino}. When properly designed,
polymer brushes in good solvent conditions have been shown to remarkably reduce
friction \cite{Klein2}. The brush structure and its properties can be controlled
by tuning the grafting density, and variation of polymer molecular weight,
temperature, and solvent quality \cite{Auroy1}. Numerous theoretical
\cite{Alexander,Milner1,Milner2,Halperin2,Biesheuvel1,Biesheuvel2,de
Gennes,Wijmans}, experimental \cite{Auroy2,Cho,Smith,de Vos1,de Vos2}, and
simulational
\cite{Grest2,Murat1,Lai1,Lai2,Lai3,Neelov1,Neelov2,Neelov3,He,Coluzza,Seidel,
Kreer,Dimitrov1,Dimitrov2} studies have examined the structure and properties of
polymer brushes. There exist, however, sometimes significant differences between
the theoretical and the experimental investigations of polymer brushes. So most
theories assume strong stretching of polymer chains in the brush, while it is
hard for an experimentalist to achieve densities high enough so as to meet this
assumption \cite{Milner3}. The polydisperse nature of the synthesized brushes
poses another significant difference. While it is practically impossible for an
experimentalist to produce a perfectly monodisperse polymer brush, to the best
of our knowledge, most of the theoretical works so far have not taken into
consideration the intrinsic polydispersity of the brushes with a realistic chain
length distribution, see, however, the early work of Milner et al. \cite{Milner4}.

The interaction between an AFM tip and a nanodesigned polymer brush bears
resemblance to the problems mentioned above. Previous works on the interaction
between a uniformly grafted polymer brush and an external object usually
neglected effects of polydispersity and high grafting density of the tethered
chains \cite{Jeon1,Jeon2,Subramanian,McCoy,Steels,Pang} even though the
properties of living polymers in the bulk have been considered by Flory
\cite{Flory}, Wheeler \cite{Wheeler}, Pfeuty \cite{Pfeuty} and others. A step
toward understanding the role of polydispersity in soft matter outdoor response
was done by the investigation of ``equilibrium polymers'' (EP) as summarized in
\cite{Wittmer2}. EPs (or, living polymers) denote polymer solutions where the
chains are dynamic objects with the unique feature of constantly fluctuating
lengths. Subject to external perturbation, concentration, or temperature change,
they are able to respond via polymerization - depolymerization reactions
allowing new thermodynamic and chemical equilibrium to be established. An
important example of such constant process of dynamic equilibrium between
polymers and their respective building units is that of surfactant molecules
forming long flexible cylindrical aggregates, the so called wormlike giant
micelles (GM) \cite{Wittmer2}, which break and recombine constantly at random
points along the sequence. EPs are intrinsically polydisperse and their
Molecular Weight distribution (MWD) in equilibrium is expected
\cite{Cates1,Flory} to follow an exponential decay with chain length.

An efficient way to create polymer brushes is the growth of living polymer
chains from active sites on a surface whereby brushes with given polydispersity
index and grafting density can be synthesized \cite{Vidal,Ruehe}. One of the
earliest works investigating theoretically the growth of polymer  chains from a
surface was carried out by Wittmer {\it et. al.}, \cite{Wittmer1}, combining
elements of diffusion-limited aggregation (DLA) \cite{Witten} with the theory of
polydisperse strongly stretched polymer brushes \cite{Alexander,Guiselin}.
Generalizing the ``needle growth'' problem \cite{Meakin}, they considered the
formation of the brush as a particular case of diffusion-limited aggregation
without branching (DLAWB).

The interaction between a polydisperse grafted layer, commonly referred to as
living polymer brush (LPB), and an AFM tip is similar to that of a polymer layer
with a particle of mesoscopic size. A theoretical discussion of the latter has
already been presented by Subramanian {\it et. al.}, \cite{Subramanian}. Their
treatment considered both dense brush (high grafting density) and 'mushroom'
(low grafting density) regime. For fixed brushes, in which the anchoring points
on the surface are immobile, they assumed that when the particle is pushed
against the grafted chains, these chains undergo pure compression and do not
splay. Under such an assumption, the force of compression (force $F$ divided
by cross-sectional area of the compressing particle $A$) per unit area, $F / A$,
does not depend on the radius of the compressing particle and is identical to
the force for a brush pressed by a flat (infinite radius) plate. In the mushroom
regime, they allowed the chains to deform and escape compression, and in this
case Subramanian {\it et. al.}, \cite{Subramanian} found that the chains do
splay to the side to avoid being compressed. A nanopatterned polydisperse
polymer brush, however, is a much more complex system as its distributed
correlation length is of the same order as the size of the system. Therefore,
the brush can evade to the side, which is impossible for the monodisperse
grafted brush, making the nanopatterned polydisperse brush effectively
``softer''.

The force measured in an AFM experiment is rarely the quantity that one is
ultimately interested in. The noteworthy exception to this statement are studies
of the chemical nature of the brush surface where the degree of sticking is the
sought-after information \cite{Magonov}. In most other cases, the measured force
$F$ first has to be converted, or ``gauged'', to yield, for example, the brush
density $\phi$. However, already for a homogeneously monodisperse polymer brush,
computing the density $\phi$ is not possible from the knowledge of the force
{\it F} alone - instead, additional information about the brush is needed
\cite{Murat2}, such as its grafting density, $\sigma_g$, and this information
frequently might not be available.

Still, for the case of monodisperse polymer brush, a relation between $F$ and
$\phi$ always exists and it also does not depend on the transversal position of
the AFM tip. All of this changes when a polydisperse polymer brush is considered
instead. Even for the same brush, identical values of the force $F$ now
correspond to different values of the density $\phi$ (or some other quantity of
interest). This hinders a deeper understanding of the polymer brush properties
via direct interpretation of AFM results.

Thus, the need to understand the response of polydisperse polymer brushes to
local external forces is both fundamental and practical. In this communication,
we first grow the polydisperse living polymer brush (LPB) {\it in situ}, monomer
by monomer, from a functionalized seed carrying polymerization initiators, that
is representative for a wider range of realistic systems. We then perform coarse
grained off-lattice Monte-Carlo (MC) simulations \cite{Milchev4,Milchev6} of the
interaction between the polydisperse living brush and a piston as representative
for the AFM tip. The simulations represent roughly $2000 - 20000$ g/mol chains
($30-260$ coarse-grained monomers per chain) over a range of grafting densities.
The high grafting densities examined in this study are achievable experimentally
by ``grafting from'' techniques such as atom transfer radical polymerization
(ATRP) \cite{von Werne,Ell}. Anticipating, we observe good agreement between
simulational and experimental force - distance relationships.

The outline of the remainder of the paper is as follows. In the next Section
\ref{sec_theory} we sketch briefly the basis of the analytical treatment of LPB
and the principal properties of living polymer brushes. In Section
\ref{Model_MC} we introduce the Monte Carlo model, employed in our study, and
elucidate the salient features of the underlying algorithm. 

The steric repulsion exerted on an impenetrable and semi-permeable wall in
compression experiments with {\it in situ} grown LPB are presented in section
\ref{sec_results}. In this section, we ask what happens when a layer of
polydisperse polymer chains, grafted to a planar substrate by special groups, is
subjected to external force (such a AFM tip). We also examine the effect of
keeping the concentration of single non-grafted monomers in the box constant by
using a semi-permeable wall so that the mean degree of polymerization in the
brush does not change when the wall is moved. In the next section
\ref{sec_comparison}, the comparison between experiment and simulation is
presented in more detail in view of our own experimental data obtained from
colloidal probe AFM measurements. We briefly discuss the physical significance
of the conversion factors for chain conformation sizes and length scales in real
application, from the simulation to experiment. Eventually, in the last section,
\ref{sec_summary}, we end this work by a brief summary of our findings.

\section{Theoretical Framework} \label{sec_theory}
\subsection{Analytical Predictions of the Distribution of Chain Length}
\label{subsec_MWD}

On a coarse-grained level, systems of living polymers are characterized by the
monomer volume fraction $\phi$, the energy difference $E$ between
saturated and unsaturated bond states, the persistence length $l_p$, and the
excluded volume size $b$ of the monomer. For chains that are long compared to
persistence length $l_p$, reversibility of the self-assembly process ensures
that the molecular weight distribution (MWD) $c(N)$ of the polymeric species is
in thermal equilibrium. Polymer brushes are usually created by means of
surface-initiated polymerization \cite{Olivier} and are characterized by
non-negligible polydispersity \cite{Knoll}. A large variety of synthetic routes
for the generation of polymer brushes includes, e.g., ionic- \cite{Advincula} ,
ring-opening- (ROP) \cite{Voccia}, atom transfer radical- (ATRP) \cite{Jankova},
and reversible addition-fragmentation chain transfer- (RAFT) \cite{Chiefari}
polymerization.

Most frequently in living polymers one observes a Flory-Schulz Molecular Weight
Distribution (MWD) distribution of chain lengths:
\begin{equation}\label{eq:length_dist}
 c(N) = (1-p_r)p_r^{N-1} = \frac{M_0}{M_n}\left( 1 - \frac{M_0}{M_n} \right )
^{N-1} ,
\end{equation}
where $c(N)$ is the fraction of chains with length $N$, $p_r \le 1$ is the
probability that a monomer has reacted, $M_0$ is the molecular weight of a
monomer, and $M_n$ - the number-averaged molecular weight. Physically,
Eq.~(\ref{eq:length_dist}) reflects a case when the reactivity of building
units is independent of macromolecular weight.

In the case of dilute chains in the bulk, it has been shown that the MWD takes
the form of Schulz-Zimm distribution \cite{Grosberg}
\begin{eqnarray}\label{eq:c_SZ}
 c(l)=\frac{\gamma^\gamma}{\Gamma(\gamma)} \left( \frac{l}{\langle L \rangle}
\right) ^{\gamma - 1} \exp\left(-\gamma \frac{l}{\langle L \rangle}
\right)\quad
\end{eqnarray}
and an mean chain length is given by
\begin{eqnarray}\label{eq:mean_L}
 \langle L \rangle \propto \phi^\alpha \exp( \delta E)\quad
\end{eqnarray}
In Eqs.~(\ref{eq:c_SZ}) - (\ref{eq:mean_L} one has \cite{Wittmer2} $\alpha
= \delta = (1+\gamma)^{-1}$ whereby $\gamma \approx 1.165$ in three dimensions.

In living polymers the semi-dilute conditions correspond to the case $L^*\ll
\left<L\right>$ and $\phi > \phi^*$, where $L^*$ and $\phi^*$ mark the mean
chain length and the density at crossover regime. Eventually, proceeding as in
the ideal case, one may recover in the semi-dilute case a simple exponential
expression for the molecular weight distribution,
\begin{eqnarray}\label{eq:mean_l_2}
c(l) = \frac{\phi}{\left<L\right>^2}{\rm
exp}\left(-\frac{l}{\left<L\right>}\right),
\end{eqnarray}
with a slightly different expression for the average polymer length
\begin{eqnarray}\label{eq:L_aver}
\langle L \rangle \propto \phi^\alpha \exp(\delta E),\quad \delta =
\frac{1}{2},\quad
\end{eqnarray}
where $\alpha = \frac{1}{2}\left(1+\frac{\gamma -1}{3\nu -1}\right) \approx
0.6$ \cite{Wittmer2}.

Turning now from living polymers in the bulk to a living polymer brush (LPB),
one retains the unique feature that the chains are dynamic objects with
constantly fluctuating lengths. Subject to external perturbation, {\em inter
alia} to pressure through a piston, they are able to respond dynamically via
polymerization - depolymerization reactions allowing new thermodynamic
equilibrium to be attained. Here we consider a typical case of living ionic
polymerization whereby chains grow from {\em initiators}, fixed on a grafting
plane, by end-monomer mediated attachment - detachment events whereby the total
number of chains remains constant.

Making use of scaling considerations \cite{Wittmer2}, one may try to predict
the power law exponents that govern LPB structure under various conditions. In
the simplest case of dilute non-overlapping living polymers (mushrooms),
tethered to a plane in a good solvent, one expects $F_{chain}(N) = \tau \ln(N)$
with the exponent $\tau = 1-\gamma_s$ where the universal 'surface' exponent
$\gamma_s \approx 0.65 < 1$. Hence, one expects to find a weakly singular $c(N)
\propto N^{-\tau} \exp(- \mu_1 N)$. Thus, in a weakly stretched polymer brush at
low grafting density $\sigma_g$, when the living polymers do not overlap
strongly, the excluded volume interactions are expected to favor longer chains
which can explore broader regions of the living polymer brush. This would lead
to a power law MWD (plus exponential cutoff). Therefore, for a self-similar
mushroom structure of the living polymer brush  with blob size $\xi(z) \propto
z$ the monomer density would scale as $\phi(z) \propto z^{-\alpha},\; \alpha =
(3\nu-1)/\nu$. The chain-end density, like the blob density, $\rho_e(z) \propto
1/\xi(z)^3 \propto z^{-\beta}$ with $\beta = 3$. Since $c(N)dN = \rho_e(z) dz$,
one readily finds a power-law distribution, $c(N) \propto N^{-\tau}$ with $\tau
= 1 + 2\nu \approx 11/5$.

In contrast, assuming that the polymer brush may be viewed as a compact layer of
concentration blobs $\xi(z) \propto \phi(z)^{-\nu/(3\nu -1)}$, within the
Alexander-de Gennes picture of a strongly stretched brush (the so called Strong
Stretching Limit - SSL), the chains are described within the Self-Consistent
Field Theory approach (SCFT) as classical trajectories, that are strongly
stretched at distances, larger than $\xi$. If scaling holds, one may then use
power-law functions to express $\phi(z) \propto z^{-\alpha},\; \rho_e(z) \propto
z^{-\beta},\; c(N) \propto N^{-\tau}$ and $z(s) \propto s^{\nu_\bot}$, as in the
case of the weak stretching limit. Assuming that in the SSL a living polymer
brush may be considered as grown by diffusion-limited aggregation (DLAWB)
without branching, one obtains then set of exponents \cite{Milchev5} as $\alpha
= 2/3,\;\beta = 2,\; \tau = 7/4,\; \nu_\bot = 3/4$ \cite{Wittmer1}. Indeed, one
can verify that these theoretical predictions agree well with earlier MC
simulation results \cite{Milchev5} and our present observations (see below) of a
dense LPB in equilibrium. Given that the DLAWB model \cite{Wittmer1} pertains
to a steady state process of brush growth under irreversible polymerization in
conditions of adiabatically slow influx of free monomers, this result appears
somewhat surprising. Even, if one could provide some arguments that the
structure of a steadily growing 'needle forest' under certain conditions may be
viewed as close to that of a LPB in dynamic equilibrium, the two cases clearly
differ, and this interesting result (whose study is beyond the scope of the
present study) certainly needs special investigation.

Regarding the force, exerted by a polymer brush on a plane at distance $D$ from
the grafting surface, the problem was first considered using the Alexander-de
Gennes \cite{Alexander} model, one assumes that the force is strictly zero for
$h < D$, (corresponding to zero osmotic pressure at the outer end of the brush).
Using a Flory type argument, one estimates a force-distance relation
\cite{Kappl}
\begin{equation}\label{eq88}
\tilde{f}(D) = \frac{k_BT h \sigma_g^{3/2}}{2 a^3 D} \left [ \left (
\frac{h}{D} \right )^{5/4}- \left ( \frac{D}{h} \right )^{7/4} \right ],
\quad h > D
\end{equation}
The first term in Eq.~(\ref{eq88}) stems from the contribution to free energy
due to confinement whereas the second term is due to the  energy of stretching.
Note that for small relative compression $\varepsilon = 1-D/h$ the force
$\tilde{f}$ scales linear in $\varepsilon$. Here $\sigma_g$ is the grafting
density, $h$ is the theoretical brush height (thickness), $k_B$ - the Boltzmann
constant, and $T$ - denotes the temperature.

Regarding the force of interaction, $f$, for the geometry of a grafted plane and
a spherical body of radius $R$ at distance $D$ between the surfaces can be
described by a slightly modified equation \cite{Helm,Butt,Curro,Vancso},
\begin{eqnarray} \label{eq_force_dist}
\frac{f}{R} = \frac{8k_BT h
\sigma_g^{3/2}}{35}\left[7\left(\frac{h}{D}\right)^{5/4}
+ 5\left(\frac{D}{h}\right)^{7/4} - 12\right].
\end{eqnarray}
where the constant term is included to ensure that the force goes to zero when
$h = D$. Eq.~(\ref{eq_force_dist}) has been used also by Plunket et al.
\cite{Plunket}.

While the theoretical results, Eqs.~(\ref{eq88}), (\ref{eq_force_dist}) have
been derived for monodisperse brushes, it is also interesting to examine the
effect of polydispersity on the effective mean height of a polymer brush, and
the ensuing force exerted by the polymer brush on a plane upon compression. As
shown by Milner, Witten and Cates \cite{Milner1}, the height of a polydisperse
brush within the SSL-SCFT is \cite{Milchev4}
\begin{equation}\label{eq_var}
 h[\sigma_g] = \sqrt{\frac{12}{\pi^2}} \int_0^{N_{max}} dn
\sqrt[3]{\sigma_g -\sigma_g(n)},
\end{equation}
where $\sigma_g(n)$ is the grafting density of chains per unit area of length
less than $n$. Therefore, one can estimate that the effective height $h(\Delta)
= h_0 (1 + \Delta / (2\bar{N}))$ (e.g., for a narrow uniform molecular weight
distribution) increases as a square root of the variance $\Delta^2$, where
$M_w/M_n = 1 + \Delta^2 / (3 \bar{N}^2)$. Accordingly, at small compression the
repulsive force of the polydisperse brush increases and is always larger than
that of a monodisperse polymer brush with chains of length equal to the mean
length of the polydisperse brush $\bar{N}$. At large compression, the two forces
should be identical \cite{MWC}. For a broad distribution $c(N)$, cf.
Eq.~(\ref{eq:c_SZ}), this effect may also be estimated, and one would expect a
quartic force law up to compressions of order $h$ \cite{MWC}.

\section{Method: Model and Simulation Aspects}
\label{Model_MC}

We use a coarse grained off-lattice bead spring model to describe the polymer
chains in our two systems: LPB and MPB. As far as for many applications in a
biological context rather short grafted chains are used \cite{Semal}, and in
order to do a reasonable comparison, we restrict ourselves to the range of
$\langle L \rangle = 28-170$ and $N = 32, 64, 128$ at various grafting densities
$\sigma_g$ in the cases of LPB and MPB, respectively.

\subsection{Living Polymer Brush (LPB)}

\begin{figure}[bht]
\vspace{0.8cm}
\includegraphics[width=8cm, height=6cm]{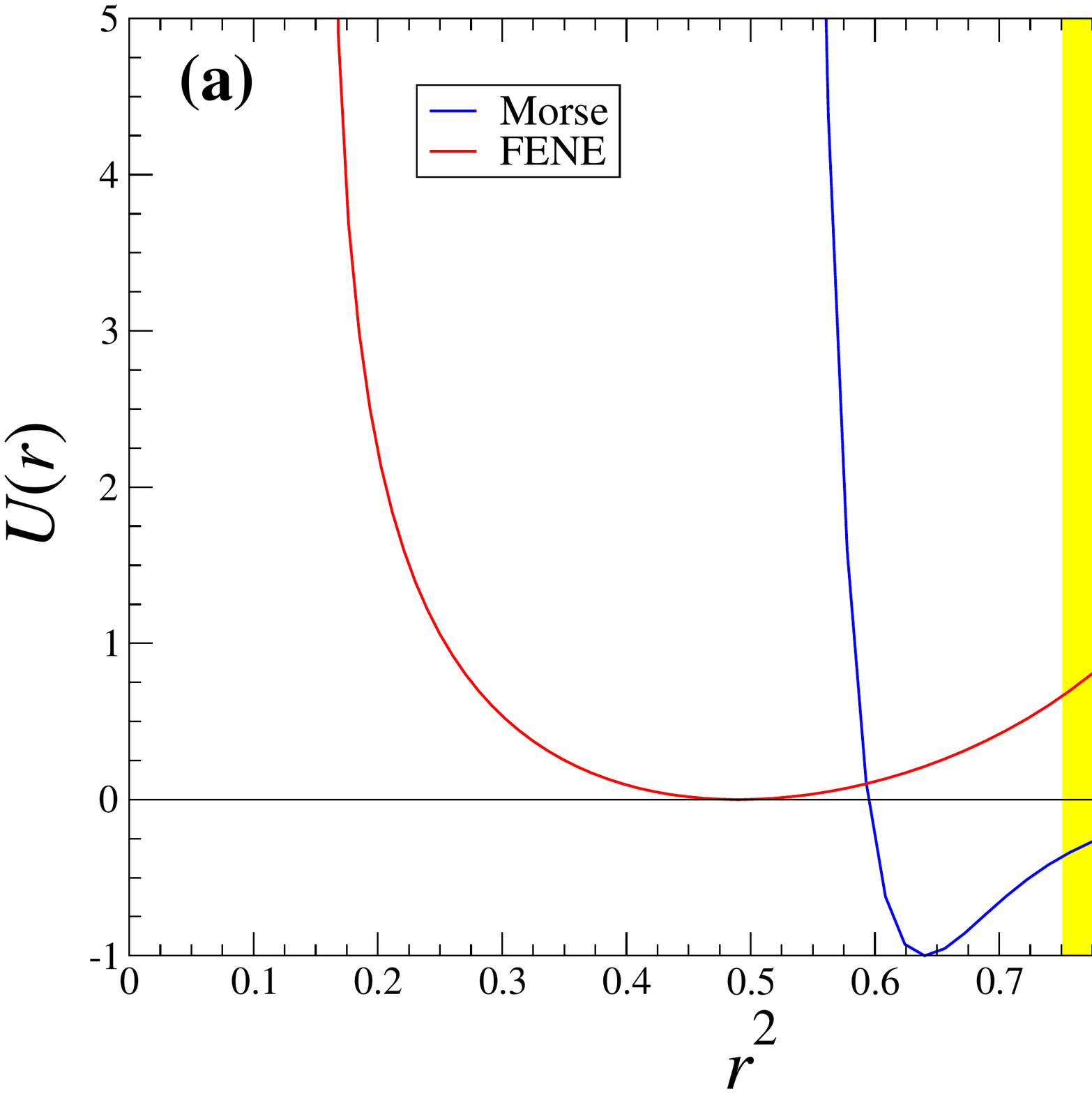}
\hspace{1cm}
\includegraphics[scale=0.4,angle=0]{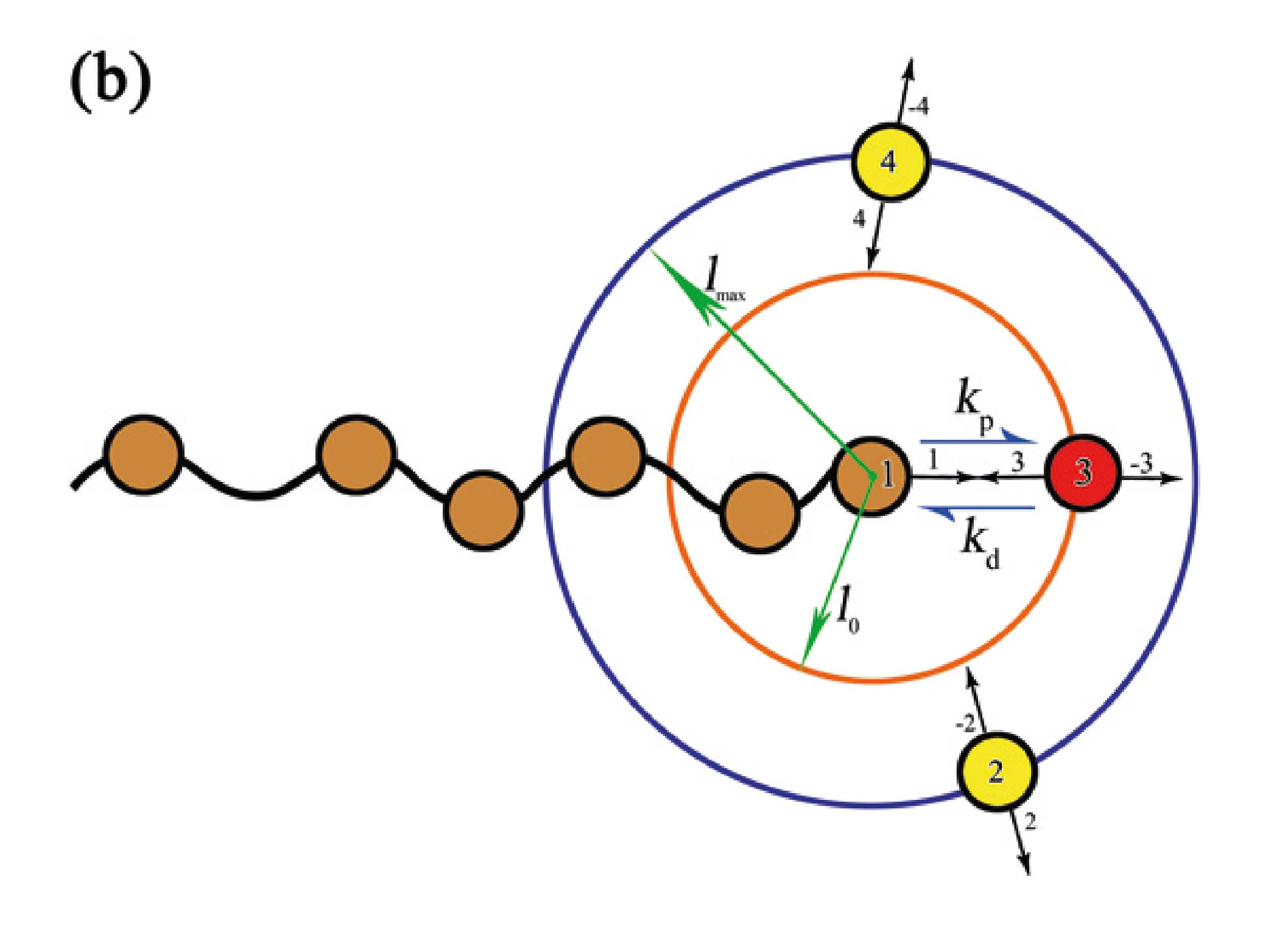}
\caption{ (a) Plots of bonded (FENE) and nonbonded (Morse) interactions used in
the present model. The shaded area denotes distances where
scission-recombination events may take place. (b) Two-dimensional projection of
the living polymerization-depolymerization process with corresponding rates of
polymerization $k_p$ and depolymerization $k_d$: chains consists of successively
connected bonds, labeled by pointers. The pointers of end-bonds point to {\it
NIL} (red particles). The breaking of a saturated bond {\it ibond} requires to
set the pointers of the two connected bonds {\it ibond} and {\it jbond} = {\it
pointer}({\it ibond}) to {\it NIL} ( yellow particles, $2,\;4$). Setting
$pointer(3)=1$ and $pointer(1)=3$ connects the end-monomer $imon=1$ and the
individual reacting monomer $jmon=3$. \label{Model}}
\end{figure}

For the off-lattice version presented here we have now harnessed a very
efficient bead-spring algorithm for polymer chains (for technical details see
Ref. \cite{Gerroff}) and this off-lattice scheme is characterized by the
bonded and non-bonded interactions shown in Figure~\ref{Model}a.

It is clear that in a system of EP where scission and recombination of bonds
constantly take place, the particular scheme of bookkeeping should be no trivial
matter \cite{Milchev2}. Since self-assembled EP chains are only transient
objects the data structure of the chains can only be based on the individual
monomers or, rather, on the saturated or unsaturated bonds of each monomer
\cite{Wittmer2}. As sketched in Figure~\ref{Model}b, in the polymerized and
depolymerized (recombination and scission) state a reacting monomer $1$ is
either randomly linked to another reacting monomer $2$ through a given bonding
potential and thus contributes to the final conformation of the polymer chain,
or non-bonded, belonging thus to the fully free single reacting monomers (yellow
particles). It is further imposed that in the state in which living
polymerization and depolymerization events take place simultaneously
(Figure~\ref{Model}b), the polymerizing monomer is located within the region of
low intermolecular energy determined by a suitable bonding potential. Note that
active chain ends do not react (recombine) with one another.

Using the assumption that no branching of chains is permitted, each bond is
considered as a {\it pointer}, originating at a given monomer and pointing to
the respective other bond with which the couple forms a nearest neighbor (brown
particles), or to {\it NIL} (red particles), if the bond is free (unsaturated).
The two possible bonds of each monomer {\it imon} are called {\it ibond} = {\it
imon} and {\it ibond} = {\it -imon}. No specific meaning (or direction) is
attached to the sign and this is merely a convenience for finding the monomer
from the bond list: {\it imon} = $\arrowvert{\it ibond}\arrowvert$. Pointers are
taken to couple independently of sign and the bonds are coupled by means of a
pointer list in a completely transitive fashion.

\begin{figure}[bht]
\includegraphics[width=8cm, height=6cm]{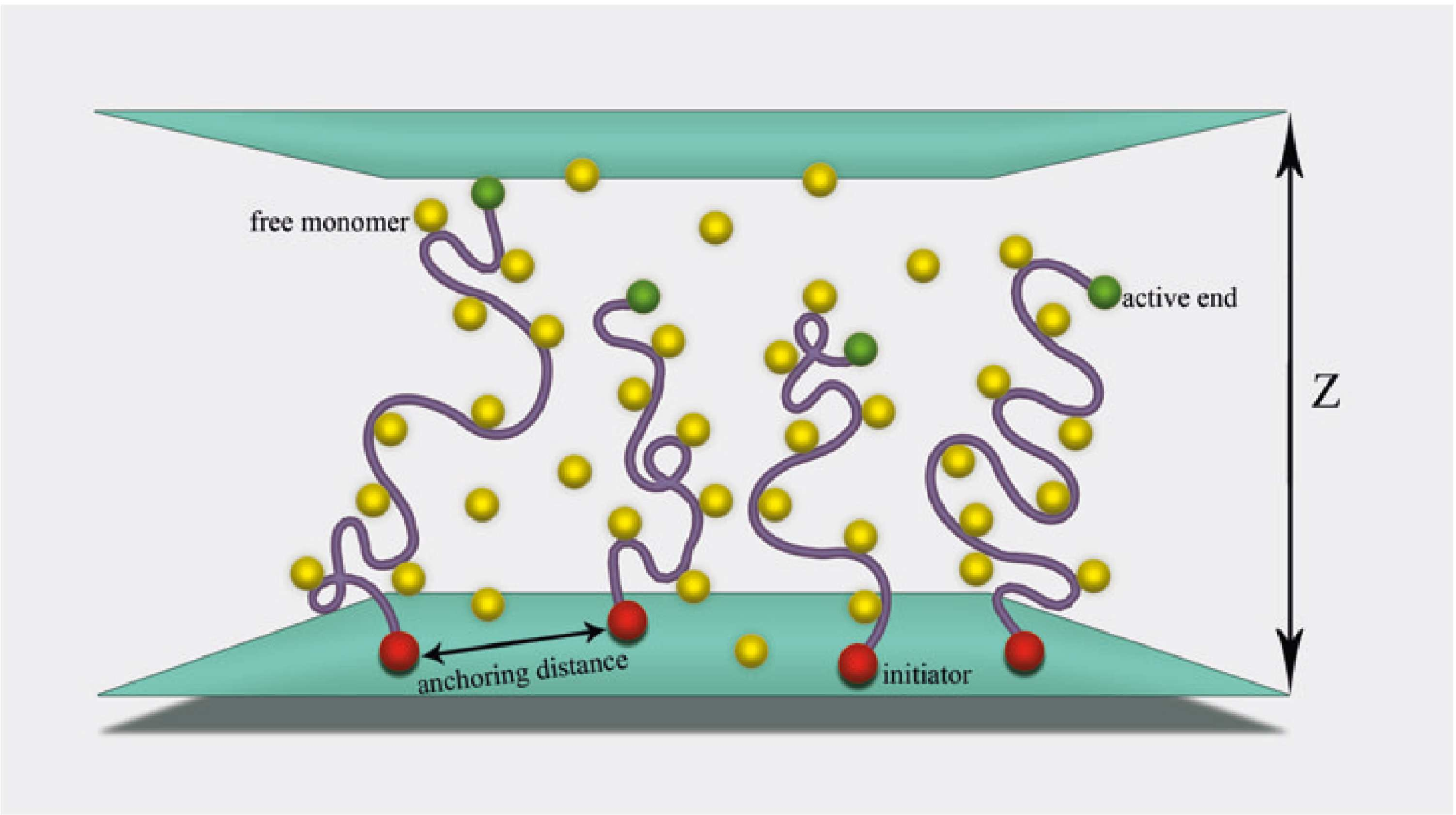}
\caption{Schematic representation of the model LPB growth between two parallel
plates. Polymer chains grow reversibly from the initiator molecules (red
spheres) fixed on the impenetrable bottom wall. They are in thermal equilibrium
with a reservoir of free monomers (gold spheres). We suppose that both the
scission energy $J$ and the activation barrier are independent of the monomer
position (along the chain contour as well as in space) and density. Desorption
events occur only at the active chain ends (green spheres) and chains are not
allowed to break along the backbone. Branching of chains is forbidden as well.
The opposite im(semi)permeable wall is a bare surface at distance $z$ from the
grafting plane. \label{scheme}}
\end{figure}

In the LPB, only two simple operations are required for polymerization or
depolymerization processes. Unsaturated bonds at chain ends point to {\it NIL}
(nowhere) and only these bonds may recombine (pair) with the unsaturated bond of
a free monomer. Therefore, the main difference with respect to EPs is that the
monomer may attach or dissociate reversibly only to / from end-monomers of
grafted LPB chains (following the model definition sketched in
Figure~\ref{scheme}) and no explicit distinction between the end-monomers,
middle monomers or free monomers is required.

\begin{figure}[bht]
\vspace{0.5cm}
\includegraphics[scale=0.3,angle=0]{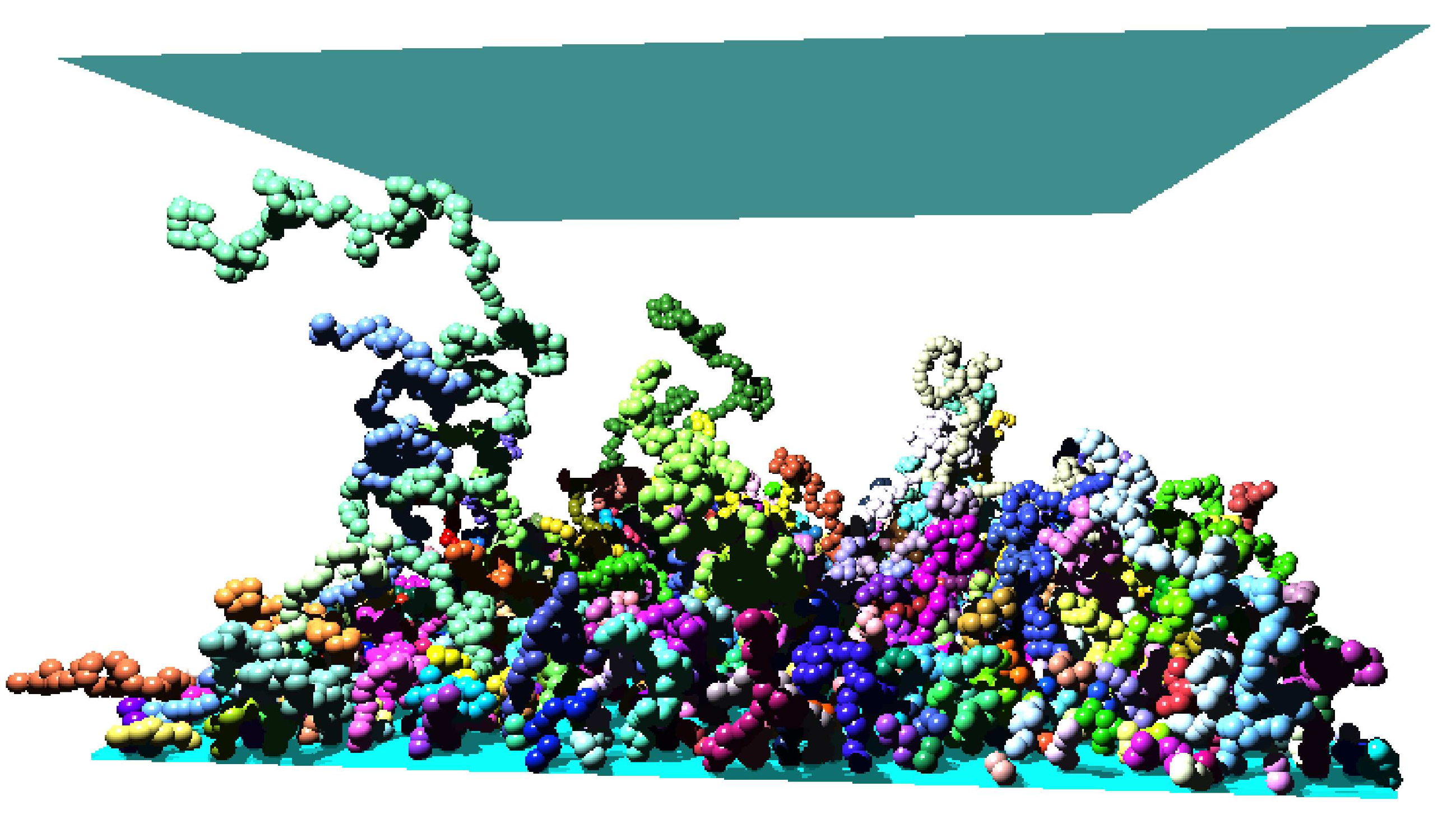}
\hspace{0.001cm}
\includegraphics[scale=0.3,angle=0]{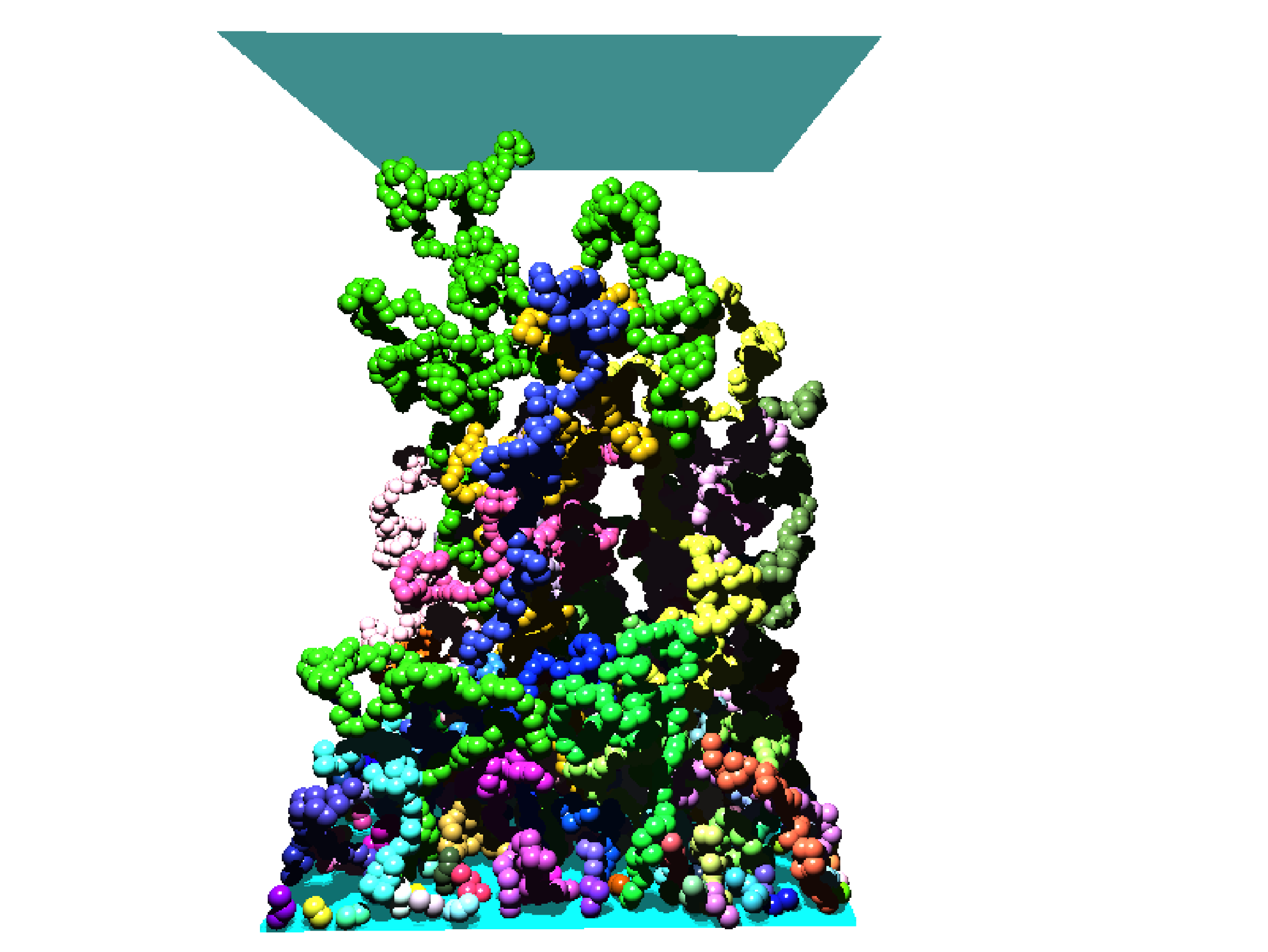}
\caption{Snapshot picture of two LPB with $\sigma_g = 1.0$, $\left<L\right> =
32$ (left part) and $\sigma_g = 1.0$, $\left<L\right> = 64$ (right part) in
their unperturbed states. The substrates are highlighted by cyan color, the
opposite walls are in green color, while monomers of polydisperse brush
belonging to short and long chains are displayed by different colors for more
distinction. The ambient free monomers are not displayed for the sake of
clarity. Only the shorter chains do sufficiently overlap and, hence, are
somewhat stretched by the density gradient.
\label{snap_LPB}}
\end{figure}

As an example, Figure~\ref{snap_LPB} shows snapshot pictures of LPB
corresponding to mean chain length $\left<L\right> = 32$ (left) and $\left
\langle L \right\rangle = 64$ (right) and grafting density $\sigma_g = 1.0$.
Each bond is described by a shifted FENE potential,
\begin{eqnarray} \label{eq:FENE}
U_{\rm FENE}(l) =
-K(l_{\rm max}-l_0)^2\ln\left[1-\left(\frac{l-l_0}{l_{\rm
max}-l_0}\right)^2\right] - J
\end{eqnarray}
where $\it J$ corresponds to the constant scission energy. Note that $U_{\rm
FENE}\left(l=l_0\right)=-J$ and $U_{\rm FENE}$ near its minimum at $l_0$ is
harmonic, with $K$ being the spring constant, and the potential diverges to
infinity both when $l\rightarrow l_{\rm max}$ and $l\rightarrow
l_{\rm min}=2l_0-l_{\rm max}$. Following Ref. \cite{Milchev2}, we choose the
parameters $l_{\rm max}-l_0=l_0-l_{\rm min}=0.3$ and $K/k_BT=20$, $T$ being the
absolute temperature. The units are such that the Boltzmann's constant $k_B=1$.
The nonbonded interactions between brush and free chain segments are described
by the Morse potential, $r$ being the distance between the beads,
\begin{eqnarray} \label{eq:Morse}
\frac{U_M\left(r\right)}{\epsilon_M}=\exp\left[-2\alpha \left(r-r_{\rm
min}\right)\right]-2\exp\left[-\alpha\left(r-r_{\rm min}\right)\right]
\end{eqnarray}
with parameters $\alpha=24$, $r_{\rm min}=0.8$, and $\epsilon_M/k_BT$ standing
for the strength of monomer-monomer interactions. In our present study we take
typically $\epsilon_M/k_BT=0.2$ which corresponds to good solvent conditions
since the $\Theta$-point of the coil-globule transition for a (dilute) solution
of polymers described by the model, Eqs. \ref{eq:FENE} and \ref{eq:Morse}, has
been estimated \cite{Milchev4} as $k_B\Theta/\epsilon_M\approx0.62$.

The model can be simulated fairly efficiently with a dynamic MC algorithm, as
described previously \cite{Milchev3,Milchev4}. The trial update involves
choosing a monomeric unit at random and attempting to displace it randomly by
displacements $\Delta x,~ \Delta y,~ \Delta z$ chosen uniformly from the
interval $-0.5\leq\Delta x,\Delta y,\Delta z\leq0.5$. The transition probability
for the attempted move is calculated from the $\delta U$ of the potential
energies before and after the move as $W = \exp(-\Delta U/k_BT)$. Moves are then
accepted according to the Metropolis criterion, if $W$ exceeds a random number
uniformly distributed in the interval $\left[0,1\right)$ and one Monte Carlo
Step (MCS) involves as many attempted moves as there are monomers in the system.
In addition, during each MCS as many bonds as there are initiators in the system
are chosen at random at the active ends of the grafted chains, and an attempt is
made to break them according to the Metropolis algorithm. Attempts are also made
to create new bonds between the end-monomers and free monomers within the
potential range of $U_M$ (i.e., a new bond with energy $U_{\rm FENE}$).

In order to keep the system in equilibrium with the ambient phase of single
free monomers and prevent the longer polymer chains from touching the top of
the container, we have used a rather low value of the bond energy $J=2.0$. The
lattice constant $s$ of the square grid of activated initiators is
taken as a rule as $s=1$ for the case of a dense brush and, as a special case
of a loose mushroom-like layer, $s=4$.

In $z-$direction the simulation box is bound by smooth (unstructured) 
impenetrable walls using the so called Weeks-Chandler-Andersen (WCA) potential,
$U_{\text{WCA}}(r)$, (i.e., by the shifted and truncated repulsive branch of the
Lennard-Jones potential), cf, Fig.~\ref{scheme},
\begin{eqnarray}
U_{\text{WCA}}(r) = \begin{cases}
4\epsilon \left[ \left( \frac{r_0}{r} \right) ^{12} - \left( \frac{r_0}{r}
\right) ^{6} \right] + \epsilon,  & \text{for}~~ r \leq 2^{1/6}r_0 \\
0,  & \text{for}~~ r > 2^{1/6}r_0
\end{cases}
\label{eq_U_WCA}
\end{eqnarray}
which prevents particles from leaving the container. In Eq.~(\ref{eq_U_WCA}) we
set $r_0 = 1.0$. Thus, while the substrate plane is fixed at $z=0$, the top
wall can be shifted in vertical direction like a piston to some desirable height
$Z_{top}$ whereby all particles at distances closer than $Z_{top} - r_0$ to it
are subject to repulsive force taken as the derivative of Eq.~(\ref{eq_U_WCA}).

Typically, boxes of size $L_x\times L_y
\times L_z$ and periodic boundary conditions in $x$ and $y$ directions have
been used in the simulations with $L_x=L_y=16, 32$ and $64\leq L_z\leq 256$
(all length in units of $l_{\max}$) for systems from $M_t=4096$ up to $32768$
monomers. Before starting to derive statistical averages, the grafted
living polymer brush is equilibrated by MC method for a period of $10^7$ MCS
(depending of the average chain length $\left<L\right>$ this period is varied)
whereupon one performs 100 measurement runs, each of length $5\times 10^7$ MCS.

In order to calculate the force, resulting from compression of the LPB, the
simulation starts with a well equilibrated {\it in situ} grown dense LPB
($\sigma_g = 1.0$) containing chains of, say,  $\left \langle L \right \rangle =
32$ and the upper wall far enough above the top of the LPB so that it is not in
contact with the brush. The wall is then gradually pushed down against the brush
with different moving rates along a line normal to the grafting surface. After
moving the wall to a specified distance from the grafting surface, the position
of the wall is held fixed and measurements are carried out.

\subsection{Homogeneous Monodisperse Polymer Brush (MPB)}

\begin{figure}[bht]
\hspace{2cm}
\includegraphics[scale=0.4,angle=0]{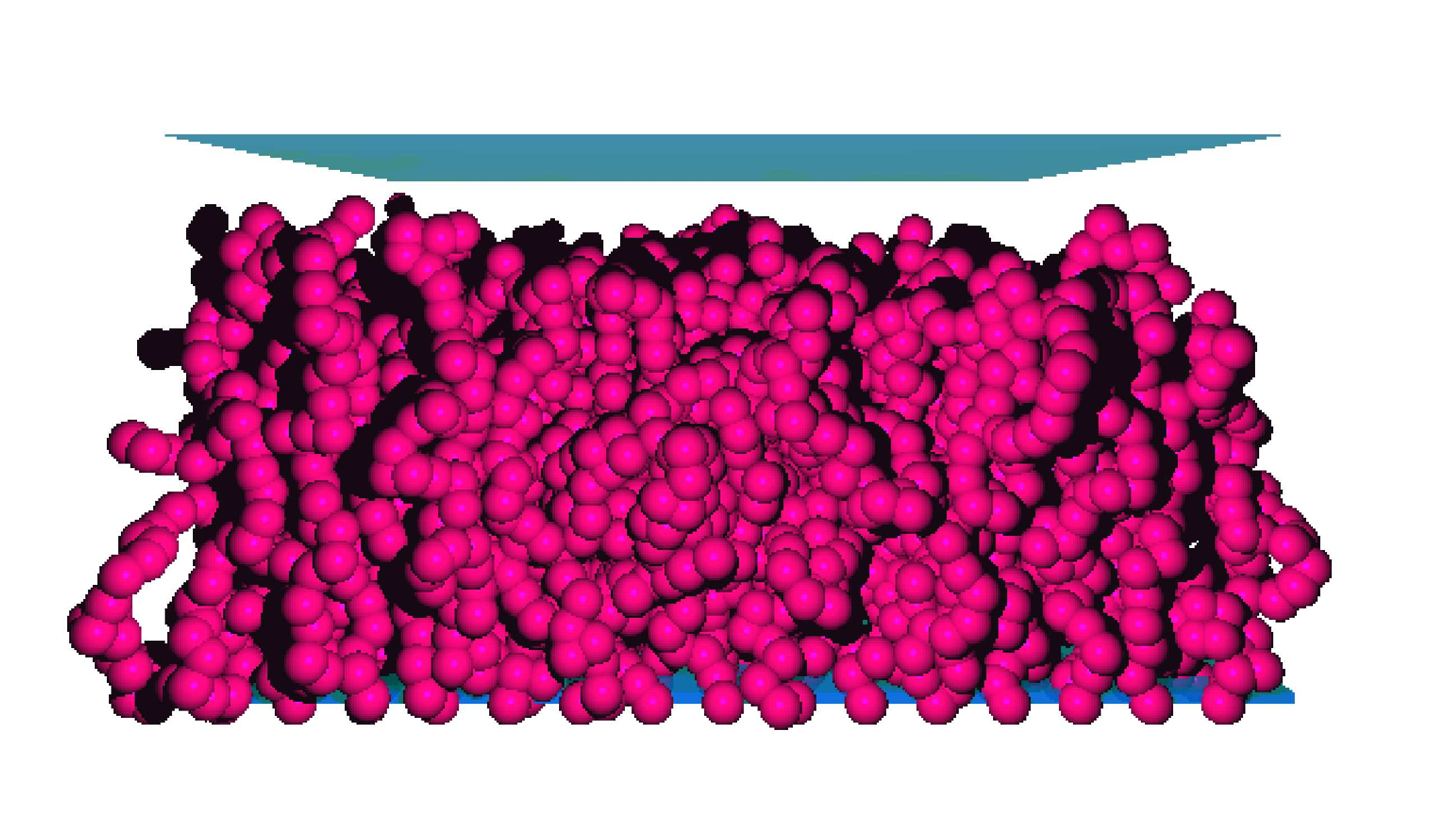} \caption{Snapshot of a
MPB with chain length $N = 32$ at grafting density $\sigma_g = 1.0$ at
equilibrium. Polymer chains are red, substrate atoms are cyan, and upper wall is
green. Solvent particles are omitted for clarity.
\label{snap_MDP}}
\end{figure}

In this case we also used the efficient off-lattice Monte Carlo method with the
first monomer of each chain being rigidly fixed to grafting sites that are
arranged regularly on a square lattice as shown in Figure~\ref{snap_MDP} . The
grafting surface is located at $z = 0.0$, and a second non-adsorbing wall is put
at $\it L_z$. The second wall is placed originally at a far enough distance from
the grafting plane such that it does not affect the chain configurations for the
chain length considered in the simulations.  We study the system of polymer
chains consisting of $N$ monomer with $N = 32,~ 64$, and $ 128$ for the highest
value of grafting density $\sigma_g = 1.0$. Periodic boundary conditions were
applied in the $xy$ direction. The brush was created in a fully stretched
configuration and was then equilibrated over times much longer than the longest
relaxation time $\tau_R$ of an single chain with excluded-volume interactions to
obtain totally independent and relaxed start configuration for compression runs.
It is well-known \cite{Gurler} that for a chain with $N$ beads and under
excluded-volume interactions, the largest (Rouse) relaxation time $\tau_R$ is
given by, in units of $N$, $\tau_R \approx 0.25N^{2.19}$ in the absence of
explicit solvent. In the monodisperse case, we have equilibrated the system for
a time $\tau = 10 \tau_R \rm MCS$.

Same as LPB system, in the MPB case, each bond is described by FENE potential
where a bond of length $l$ has a maximum at $l_{max} = 1.0$ (Eq. \ref{eq:FENE})
and the nonbonded interaction between effective monomers is described by a
Morse-type potential (Eq. \ref{eq:Morse}).

\section{Experimental studies} \label{sec_exp}

{\bf Materials.} {\it N}-Isopropylacrylamide (NIPAAm, Aldrich, 97\%) was
purified by passing through an inhibitor removal column using a mixture of
dichloromethane and hexane (v/v $\approx 1:1$) as the solvent and then
recrystallized twice from a toluene/hexane solution (50\% v/v) and dried under
vacuum prior to use. Copper(I) bromide (CuBr, Aldrich, 98\%) was purified by
stirring in glacial acetic acid, filtering, and washing with ethanol three
times, followed by drying in vacuum at room temperature overnight. Copper(II)
bromide (Sigma-Aldrich, $\geq$ 99\%),
$N,N,N',N'',N''$-pentamethyldiethylenetriamine (PMDETA; 98\%, Acros Organics),
ethyl 2-bromoisobutyrate (E2Br-iB, 98\%, Aldrich),
(3-Aminopropyl)trimethoxysilane (APTMOS, 99\%, Acros), bromoisobutyryl bromide
(BIBB, 99\%,
Aldrich), and anhydrous toluene (Merck) were used as received. Milli-Q
water (with a minimum resistivity of 18.2 M$\Omega$-cm) was obtained from a
Millipore Direct-Q 5 ultrapure water system. THF for
reactions and washings were dried by sodium before use. Double-sided polished
silicon wafers (P-doped, (100)-oriented, 10-20 $\Omega$-cm resistivity,
0.56-mm thickness) were supplied by University Wafer Company (Boston, MA) and
cut
into $5\times5 \rm mm^2$ pieces using a Micro Ace Series 3 dicer (Loadpoint Ltd,
England). Ultra-high-purity-grade argon was used in this study.

{\bf Initiator Synthesis.} The ATRP initiator,
2-bromo-2-methyl-$N$-3-[(trimethoxysilyl)propyl]propanamide (BrTMOS), was
synthesized using
a procedure modified from literature \cite{Sun}: To a stirred solution of APTMOS
(1.79 g, 10 mmol)
and triethylamine (1.01 g, 10 mmol) in 50 mL of dried THF, BIBB (3.45 g, 15
mmol) was added drop-wise at 0 $^\circ$C for 2 h under argon. The
reaction was heated to room temperature and kept for 12 h with stirring and
under argon protection. The precipitate was filtered off using a frit funnel.
The product was a yellowish oil after the removal of the solvent. The product
was redissolved with $\rm CH_2Cl_2$ (20 mL) and washed with 0.01 N HCl
(2$\times$ 20 mL) and cold water (2$\times$ 20 mL), respectively. The organic
phase was dried with anhydrous $\rm CaCl_2$. After the removal of the solvent,
the final product was a colorless oil with a yield of 90.5\%. $^1$H NMR (300
MHz, $\rm CHCl_3$): 6.91 (s, 1H, NH), 3.49 (s, 9H, $\rm SiOCH_3$), 3.25 (t, 2H,
$\rm CH_2N$), 1.95 (s, 6H, $\rm CH_3$), 1.68 (m, $\rm 2H, CH_2$), 0.67 (t,
2H, $\rm SiCH_2$). $^{13}$C NMR (600 MHz, $\rm CHCl_3$): 171.98, 62.57, 50.29,
42.62, 32.44, 22.52, 7.64.

{\bf Surface-initiated atom transfer radical polymerization (SI-ATRP).} Silicon
wafers were sonicated for $5$ min in ethanol and water, activated for $30$
min at $150$ $^\circ$C in piranha solution ($\rm H_2O_2$ (30 wt\% in
$\rm H_2O)/H_2SO_4$ (98 wt\%), v/v $\approx 3:7$) {(\bf CAUTION}: {\it piranha
solutions
are strongly oxidizing and should not be allowed to contact organic solvent.})
,thoroughly rinsed with water and
ethanol, dried in a stream of argon. A self-assembled monolayer (SAM) of the
ATRP initiator was attached to the silicon wafer by immersion in a $10$ mM
solution of BrTMOS in dry toluene at room temperature overnight. Next, the
wafers were rinsed with anhydrous toluene, sonicated for $1$ min in acetone and
in a water/$tert$-butanol mixture (v/v $\approx 1:1$), rinsed with water, dried
in a
stream of argon and transferred to the appropriate reactors for the
polymerizations.

SI-ATRP of NIPAAm monomer was carried out at room temperature from a monolayer
containing the SAM-Br initiator as follows: A mixture of MeOH/$\rm H_2O$ (v/v
$\approx 1:1$)
 was degassed through four freeze-pump-thaw cycles before being introduced
into the controlled atmosphere glove box. NIPAAm ($5.0$ g, $44$ mmol), CuBr
($40.0$ mg, $0.278$ mmol), $\rm CuBr_2$ ($\rm 14~mg$, $\rm 0.063~mmol$),
and PMDETA ($175$ $\rm \mu L$, $0.835$ mmol) were
dissolved in $30$ mL of MeOH/$\rm H_2O$ inside the glove box. The solution was
then transferred via a cannula into a vial containing silicone wafer with
uniform initiator SAM and sacrificial initiator E2Br-iB ($10$ $\rm\mu L$, $0.07$
mmol), and then vial was sealed and kept at room temperature for polymerization.
After polymerization for a times ranging from $\rm 30~min$ to $\rm 120~min$, the
substrate were removed from the vial and rinsed with copious amount of DI water,
followed by sonication in EtOH and then $\rm H_2O$. After drying with a stream
of argon, the samples were stored under Ultra-high-purity-grade argon.

{\bf AFM measurements.} Force measurements were carried out with a NanoScope
IIIa multimode atomic force microscope (Digital Instruments, Veeco-Bruker, Santa
Barbara, CA, USA) equipped with a standard liquid cell on SI-ATRP grown PNIPAAm
films in a liquid environment filled with deionized water. One micrometer in
diameter $\rm SiO_2$ colloidal probe (Novascan Technologies, Inc., Ames, CA,
USA) with a spring constant of $\rm 0.050~\pm~0.003~Nm^{-1}$ (determined using
the thermal tune method) and a diameter of $\rm 1.00~\pm~0.006~\mu m$
(determined by a scanning a tip array) was used in the experiments. PNIPAAm
coated wafer was placed on liquid cell and degassed DI water was injected into
the cell and then the film was allowed to equilibrate for $\rm 1~hr$. The AFM
cantilever was then set on top of the sample and started to move toward the
sample surface with Z-ramp size of $\rm 2~nm$ and a dwell time of $\rm 100~ms$
per step. Cantilever deflection was recorded and averaged at each interval until
tip-sample forces caused a deflection by $\rm 10~nm$, after which the cantilever
was withdrawn from the surface in the same manner.

\section{SIMULATION RESULTS} \label{sec_results}
\begin{figure}[bht]
\hspace{-2.0cm} \vspace{.0cm}
\includegraphics[angle=0,width=0.43\textwidth,  origin =
c1,]{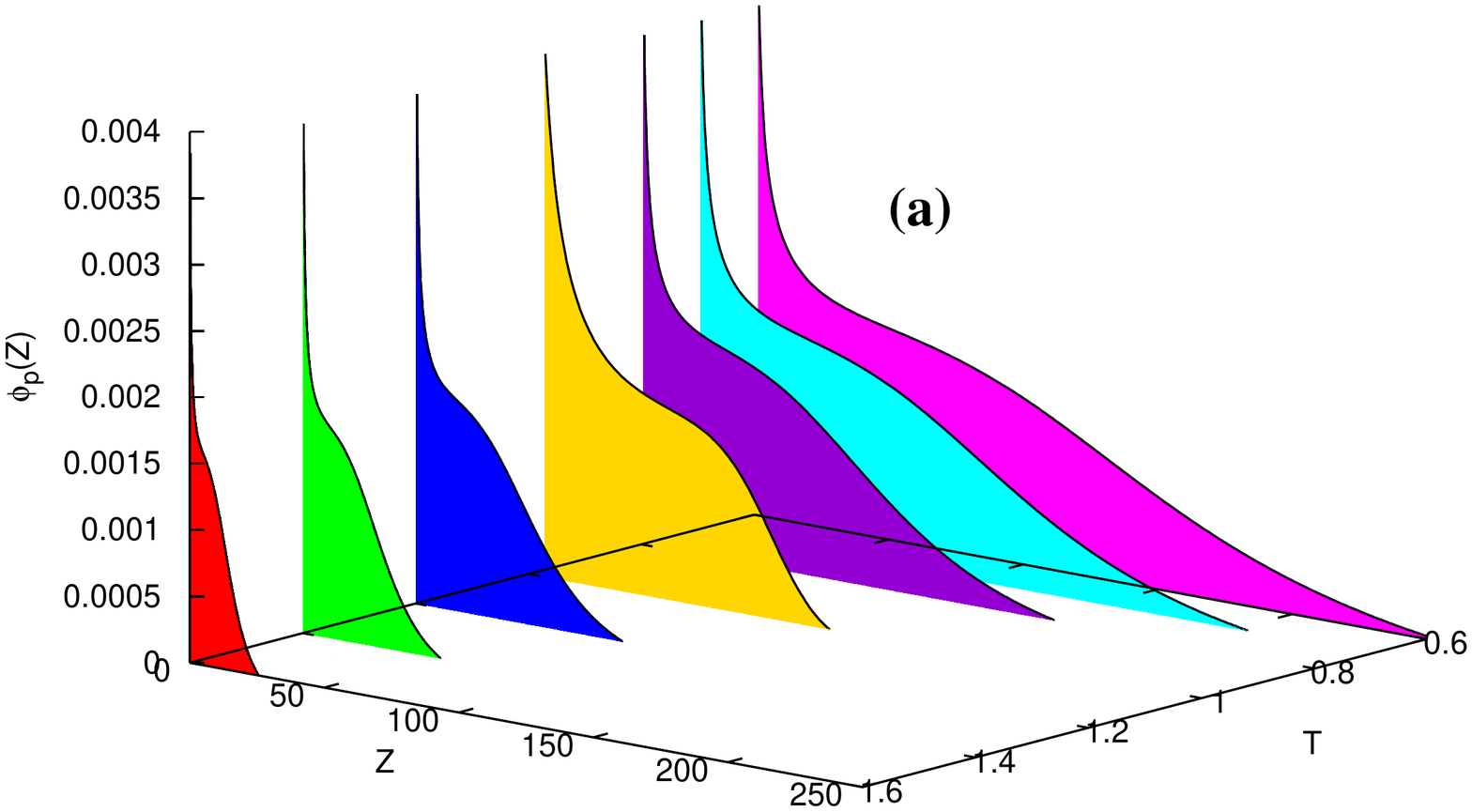}
\hspace{0.5cm}
\includegraphics[totalheight = 0.29\textwidth, angle =0]{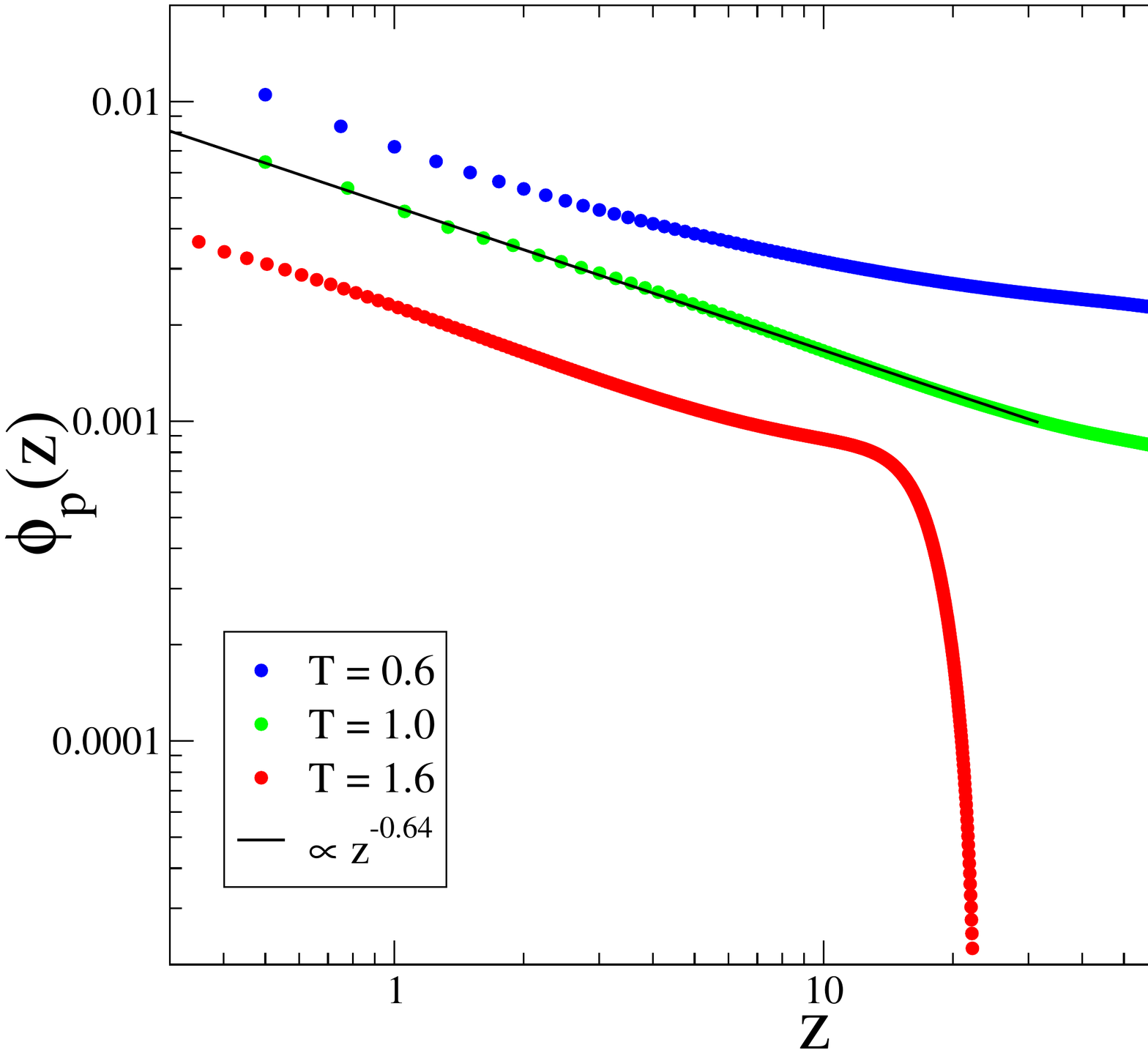}
 
\caption{(a) Density profile of LPB, $\phi_p(z)$, at different temperatures (in
units of $k_{B}T$) for LPB with total density of the system $\phi_t = 0.5$ at
$\sigma_g = 1.0$. (b) Double logarithmic plot of density profiles at different
temperatures, confirming the scaling behavior $\phi(z) \propto z^{-\alpha}$.
\label{Histo_Dynamics}}
\end{figure}

%\subsection{Polymerization / Depolymerization Kinetics of {\it in situ} Grown
%LPB}\label{subsec_quench}
As far as our main concern in this work is the comparison of experimentally
measured compression force of a LPB with data, we mention very briefly some
related results obtained from our Monte Carlo simulation. An interesting
question thereby is to what extent the structure of the LPB meets theoretical
predictions, mentioned in Section \ref{sec_theory}. Therefore, in
Figure~\ref{Histo_Dynamics}(a) we present the observed density profiles of
grafted monomers, $\phi(z)$, in an equilibrated LPB system at various
temperatures after a $T$-quench from an initial equilibrium state at $k_BT =
1.0$. It should be noted that all profiles are normalized to unity as
$\int_0^z\phi (z)dz = 1$. Evidently, in logarithmic coordinates,
Figure~\ref{Histo_Dynamics}b, one observes a power law decay of the density
profiles, $\phi(z) \propto z^{-\alpha}$, manifested as a straight line, whereby
the observed power $\alpha \approx 0.64$ at $T = 1.0$ is shown to be in good
agreement the theoretically expected one $\alpha \approx 2/3$ \cite{Milchev5}.

\begin{figure}[bht]
%\hspace{2cm}
\includegraphics[scale=0.35]{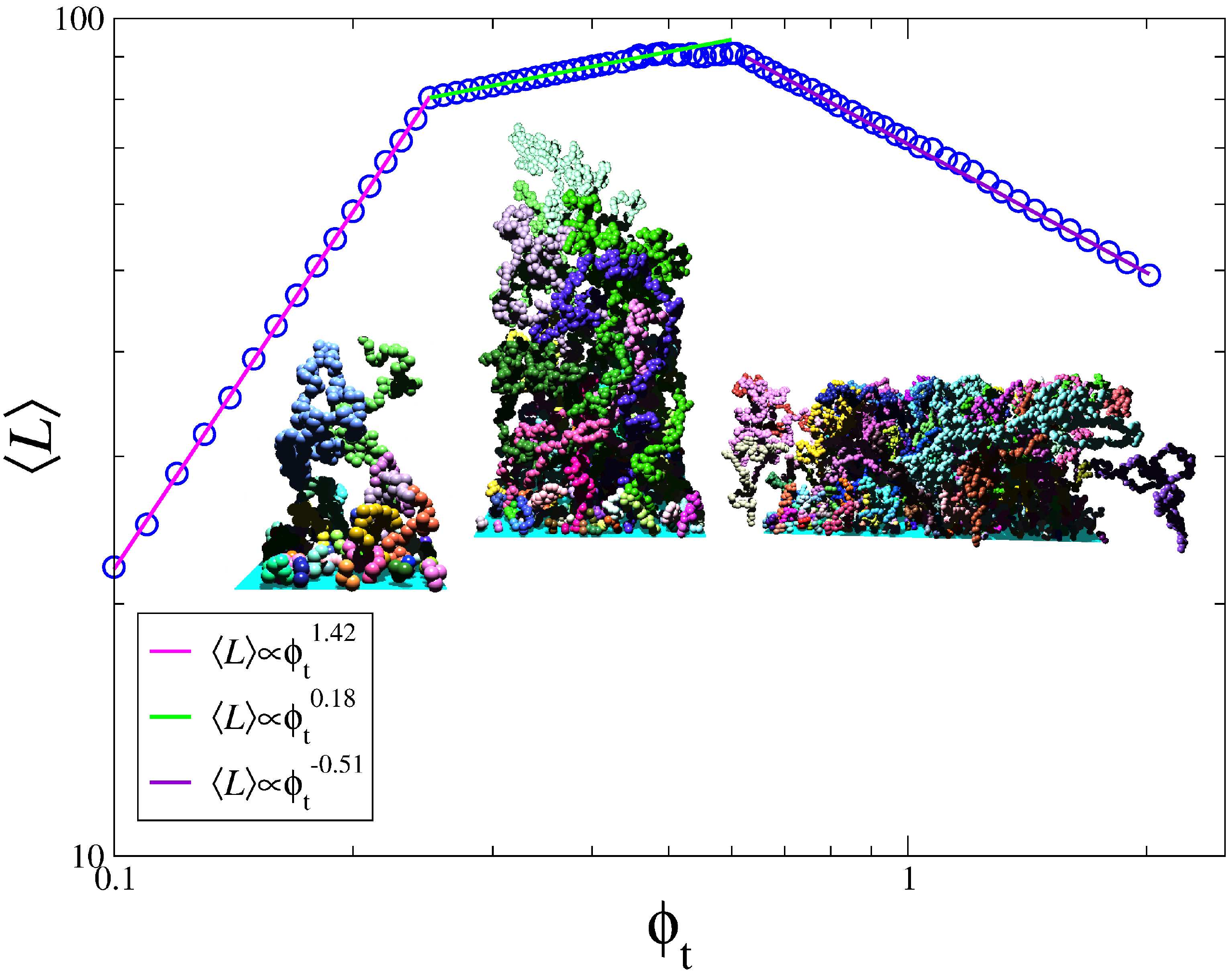}
\caption{The average chain length $\langle L\rangle$ for a wide range of total
monomer concentration $\phi_t$ confirming the scaling Eqs. \ref{eq:mean_L} and
\ref{eq:L_aver}. The three slopes correspond to dilute, semi-dilute, and
concentrated regimes at $\sigma_g =1.0$. The observed slopes are $1.42,\; 0.18,$
and $-0.51$ respectively. Sample snapshots belonging to each regimes are
inserted in the plot. \label{LvsPhi}}
\end{figure}

The data shown in Fig.~\ref{Histo_Dynamics} corresponds to equilibrated LPBs
whereby the total monomer concentration in the container has been kept constant.
We observe the same value of the exponent $\alpha$ when the temperature is
further increased.  Small deviations from the expected scaling behavior are
found only during quenching of LPB chains to lower temperatures,
Figure~\ref{Histo_Dynamics}(b), for the higher $z$. We believe that this is
probably due to single rather long chains which get repelled by the ceiling of
the simulation container and bend backwards, increasing the local monomer
density under the upper plane of the box. The reader may get impression about
such effects from the snapshots shown in Fig.~\ref{LvsPhi} where we display the
variation of the average chain length $\langle L \rangle$ in a LPB with the
total monomer concentration $\phi_t$.

The measured variation of the mean chain length $\langle L\rangle$ versus total
monomer concentration, along with visual evidence from snapshots taken in the
course of compositional variation, Figure~\ref{LvsPhi}, suggest the existence of
several distinct regimes in terms of $\phi_t$. One may conclude from
Figure~\ref{LvsPhi} that the mean chain length of a LPB in a wide range of
concentration is governed by a power-law relationships, confirming the scaling
Eqs.~(\ref{eq:mean_L}) and (\ref{eq:L_aver}) with exponents that differ from the
predicted values for living polymers in bulk. Interestingly, at rather high
concentrations the mean length $\langle L\rangle$ {\em decreases} with further
growth of $\phi_t$, i.e., the LPB gets more compact as the longest chains are
hindered in their growth due to the finite size of the simulation box. Thus, the
polymerization process is shifted and this affects increasingly the many short
chains living in the bottom. Clearly, more comprehensive investigation and
increased computational efforts are needed to elucidate this aspect of the LPB
behavior which we postpone for a separate publication.

\subsection{Compression of {\it in situ} Grown LPB}

In this section we examine the interaction between a LPB chains immersed in a
good solvent and a test surface using coarse-grained off-lattice Monte-Carlo
simulations. Our purpose is to obtain force-displacement curves for the
penetration of such an AFM tip (test surface) into a polydisperse chains of LPB
as a function of layer characteristics. To this end, the test surface is
\begin{figure}[htb]
\hspace{-2.0cm}
 \includegraphics[width=0.35\textwidth]{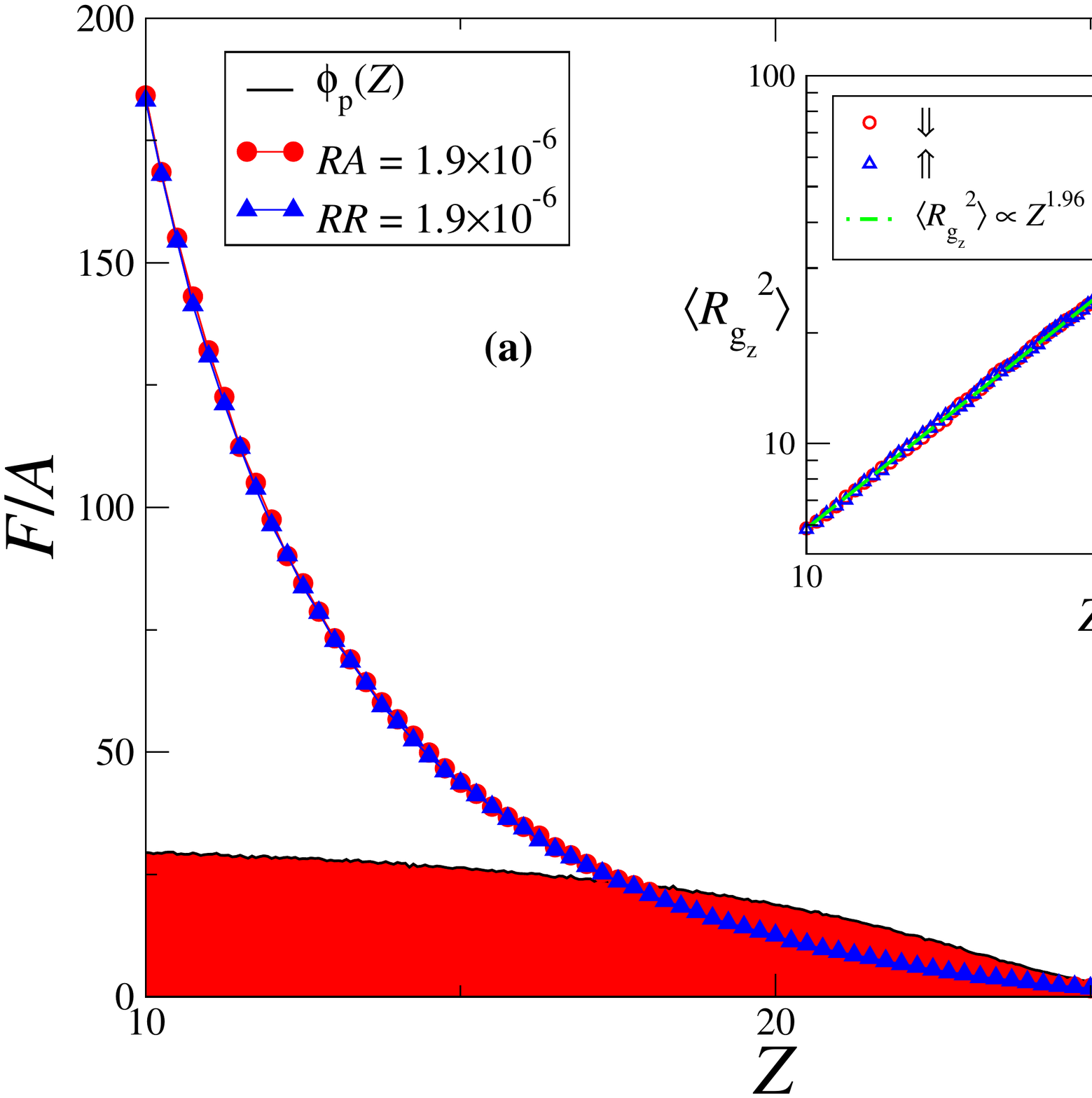}
\hspace{0.3cm}
 \includegraphics[totalheight=0.5\textwidth,origin=br,angle=0]{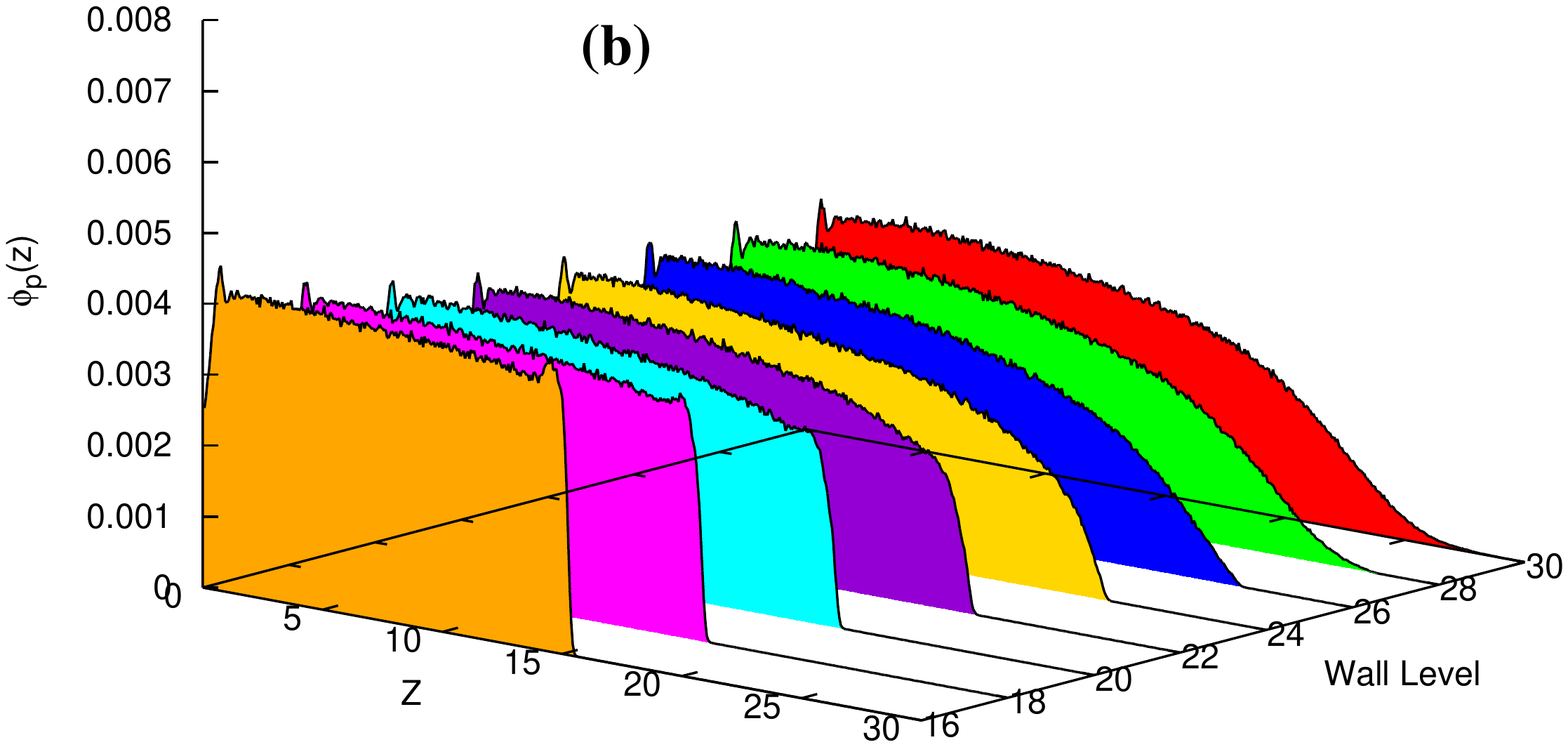}

\caption{(a) Force-displacement curves for relaxed MPB with grafting density
$\sigma_g = 1.0$ and chain length $N = 32$ during the loading $\Downarrow$ and
unloading $\Uparrow$ process at $k_{B}T = 1.0$. The rates of approaching and
retraction as unit of ${\rm MCS}^{-1}$ are given as parameter. The inset shows a
plot of the mean squared gyration radius component $\left \langle R_{g_z}^2
\right \rangle$ in direction perpendicular to the grafting surface vs height of
the mobile compressing surface (repulsive wall) during loading and unloading.
(b) Density profile of MPB, $\phi_p(z)$, for different wall positions in the
loading experiment. \label{Force_Mono_Hyster}}
\end{figure}
represented either as an impenetrable or as semi-permeable surface, consisting
of a long cylinder of cross-section area $A$. The interaction between a LPB and
a AFM tip can easily be simulated by a slight variation of the algorithms used
to study forces between two test surfaces in which one of them bears an
end-grafted polymer brush. In the simulations results given below, we assumed a
repulsive force between the monomers and the surface \cite{Murat3}, taken as a
Weeks-Chandler-Andersen potential, i.e. as the repulsive (and shifted) branch of
a Lennard-Jones potential. One might also use a Lennard-Jones with an attractive
short-range components \cite{Grest3,Grest4}. Both of these forces gave
qualitatively similar results.

For the sake of comparison, we also simulated an equilibrated monodisperse
polymer brush (MPB) containing chains of equal length $N = 32$ at high grafting
density $\sigma_g = 1.0$. Originally, the mobile wall is far enough so that it
is not in contact with the brush. To simulate the loading and unloading
processes of wall, we move it downwards at rate $RA$, and then up at retraction
rate $RR$, measured as distance per time units of MCSteps.

Figure~\ref{Force_Mono_Hyster}(a) shows the results of the force-displacement
curve for an equilibrated high density MPB at vanishing rates $RA = RR = 1.9
\times 10^{-6}$. The force $F$ is normalized by the area $A$ of contact and
describes thus the pressure on the wall. As shown in the main panel of
Figure~\ref{Force_Mono_Hyster}(a), during the loading from $z_i=30$ to $z_f=10$,
the tip goes into the sample down to a depth $\delta = z_f-z_i$, causing
deformation of the brush. During the wall retraction, it goes back from $z_f=10$
to $z_i=30$. Since the rate of approaching and retraction at well equilibrated
conditions are negligible, the MPB behaves as an elastic material and regains
step by step its own shape, exerting equal pressure on the wall. The $F$ vs $z$
curves for the approaching / retracting wall lie on top of each other and are
virtually indistinguishable. Therefore, during such an ``infinitely`` slow
loading and unloading one observes no hysteresis in the force - height
relationship. In the inset to Figure~\ref{Force_Mono_Hyster}(a) we plot the mean
squared gyration radius component $\left \langle R_{g_z}^2 \right \rangle $ in
direction perpendicular to the grafting surface as function of the wall
position. Evidently, in logarithmic coordinates these transients appear as
straight lines, suggesting that the height of the perturbed MPB changes by a
power law, $\left \langle R_{g_z}^2\right \rangle \propto z^\alpha$, whereby the
observed power $\alpha \approx 1.96$, i.e. $h \propto z$ for $z < 30$.

\begin{figure}[htb]
\vspace{0.8cm}
\includegraphics[scale=0.24,angle=0]{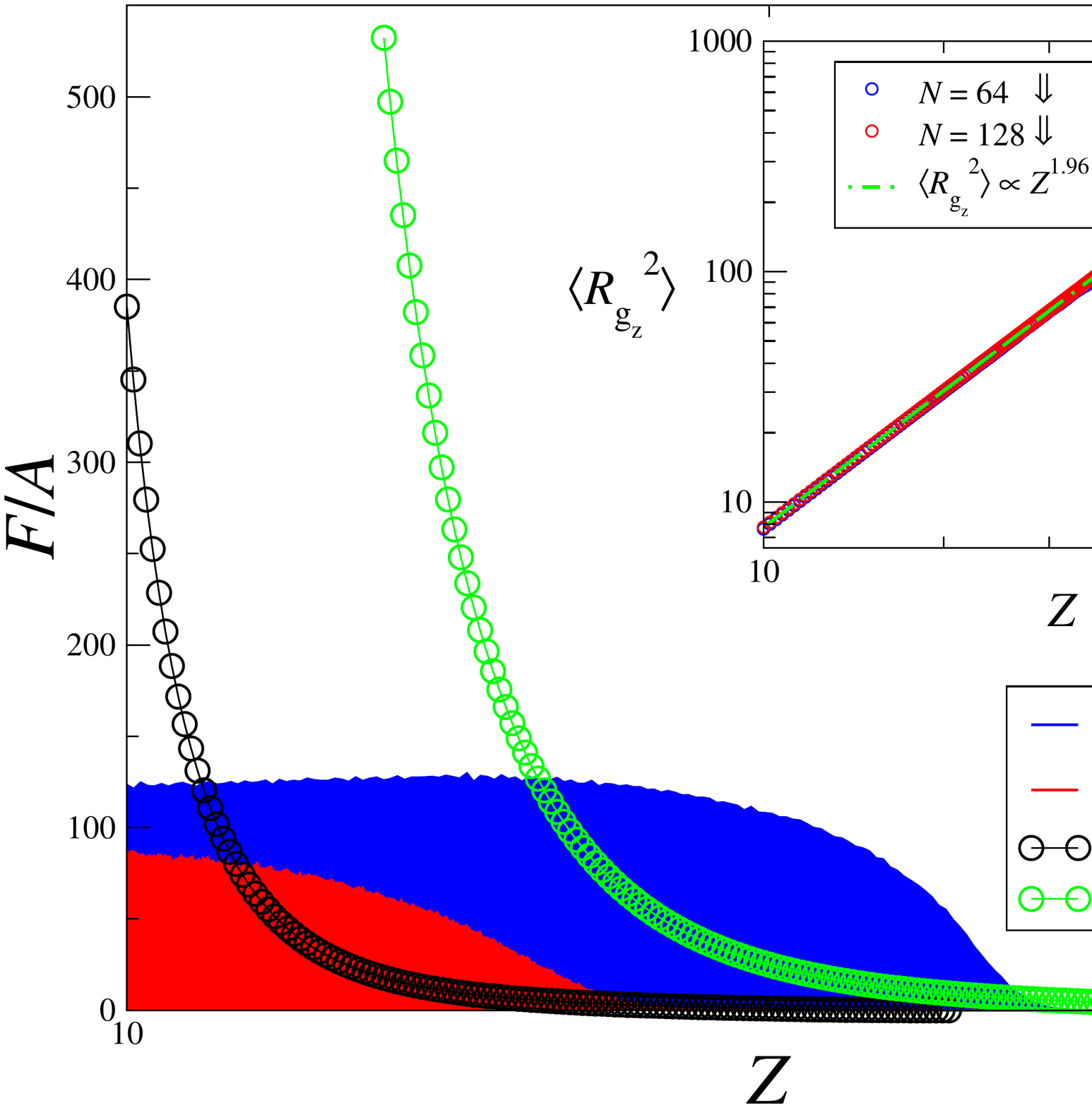}
\caption{Comparison between force-displacement curves of two relaxed MPB with
grafting density $\sigma_g = 1.0$ and chain length $N = 64$ and $N = 128$ in
the loading experiment $\Downarrow$. The density profile (shaded area) is
included for comparison. The inset shows the same trend
for $\left<R_{g_z}^2\right>$ as $N = 32$.
\label{Force_64_128}}
\end{figure}

In Figure~\ref{Force_Mono_Hyster}(b), we show the monomer distribution of the
MPB along the $z$-direction at different piston positions (wall level) during
the loading experiments. As can be seen from Figure~\ref{Force_Mono_Hyster}(b),
the histograms follow indeed the parabolic characteristics of ''dead``
monodisperse brushes \cite{Winehold} in the beginning of the process (nearly
unperturbed state) while at stronger compression, for $z\leq 22$, the histograms
indicate a flat uniform distribution of matter in the compressed MPB.
Eventually, one ends up with a rather sharply peaked $\phi_p(z)$ (not included
for the sake of clarity) near the grafting surface, characterizing a ``proximal
zone'' with density oscillations typical for all fluids ``packed'' in the
vicinity of a hard wall.

In the main panel of Figure~\ref{Force_64_128} we compare the pressure of
grafted monomers along direction $z$, normal to the grafting surface for two
MPBs with chain lengths $N = 64$ and $N = 128$, both at highest grafting
density, i.e., $\sigma_g = 1.0$. The density profiles are included for the sake
of comparison. It becomes evident from Figure~\ref{Force_64_128} that both
simulations demonstrate a qualitatively similar behavior to that of picture
Figure~\ref{Force_Mono_Hyster}. In the inset to Figure~\ref{Force_64_128} one
may see the same scaling of mean squared gyration radius component $\left
\langle R_{g_z}^2\right \rangle$ in direction perpendicular to the grafting
surface as in Figure~\ref{Force_Mono_Hyster}(a), whereby the force-displacement
curves again collapse on a single master curve. This behavior is characteristic
for large tip AFM experiments \cite{Subramanian}. Because of the large grafting
density and the excluded-volume interactions, the chains are strongly stretched
and resemble rods with small probability for the chains to escape compression by
moving away from the test surface.

\begin{figure}[htb]
\vspace{0.8cm}
\includegraphics[scale=0.3,angle=0]{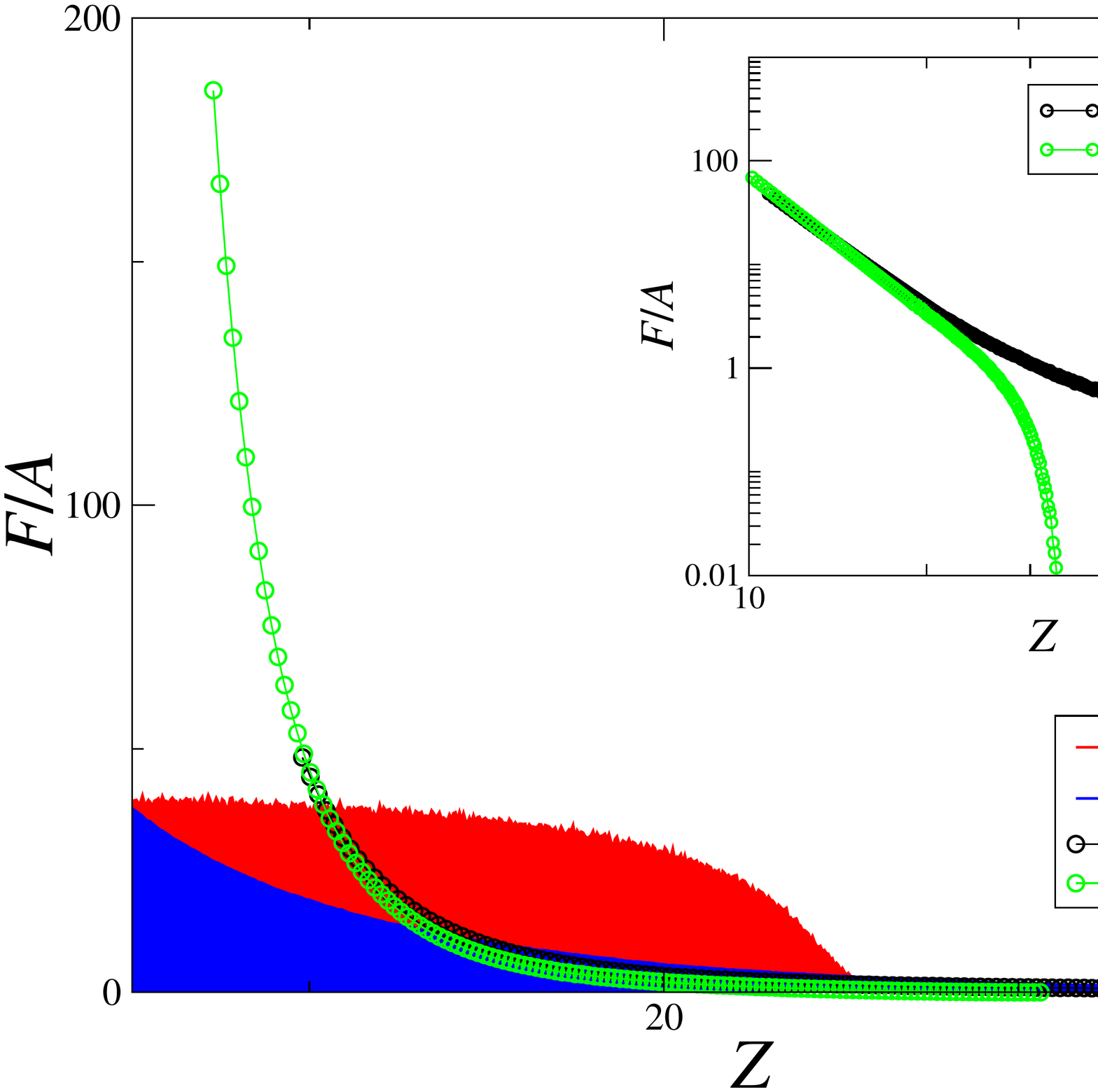}
\caption{Force-displacement curves of relaxed LPB and MPB in loading
experiment for $\left \langle L \right \rangle = 32$ (LPB) and $N = 32$ (MPB) at
$\sigma_g = 1.0$. The rate of wall displacement is $RA = 1.9\times
10^{-6}$ for both LPB and MPB. \label{Force_Mono_EP}}
\end{figure}

In Figure~\ref{Force_Mono_EP}, a force-displacement approach curve obtained for
LPB system with $\left \langle L \right \rangle = 32$ is compared to the
equivalent MPB with length $N=32$ of the grafted chains. For both cases the data
corresponds to polymer brushes at their highest grafting density, i.e.,
$\sigma_g=1.0$. For better interpretation, the resulting brush profiles are also
included in the main panel of Figure~\ref{Force_Mono_EP} by shaded area.
It is common to define the layer thickness $\langle z \rangle$ of a MPB as the
first moment of the monomer density profile $\phi_p(z)$, i.e., the average
distance of a chain monomer from the surface

\begin{eqnarray}
\langle z \rangle = \frac{2\int z\phi_p(z)dz}{\int \phi_p(z)dz}
\end{eqnarray}
The equilibrium profile of such a monodisperse ``dead'' brush is more compact
with a steeper slope near $z \approx \langle z \rangle$, than that of the
analogous LPB system (as shown in Figure~\ref{Force_Mono_EP}) which goes to far
larger $z$. In agreement with the theoretical predictions \cite{Milner4}, cf.,
Section \ref{sec_theory}, it is seen from Figure~\ref{Force_Mono_EP} that the
weak compression force of the LPB exceeds significantly the one due to MPB.

At large compression ($z < \langle z \rangle = 22.75$) or small separations, of
course, the two force-displacement relations are indistinguishable. The pressure
in this limit is exerted by the uniform density of the compressed brush
suggesting that the repulsive force is well approximated by the corresponding
osmotic pressure of a non-grafted semidilute solution. Hence the two interaction
forces become identical when the MPB height $\langle z \rangle = 22.75$ is much
less than the average chain length $\langle L \rangle = 32$ of the corresponding
LPB case.

For small compressions ($z \approx \langle z \rangle  = 22.75$), the
force-displacement curve of the MPB vanishes abruptly in contrast to the
equivalent LPB case (see inset to Figure~\ref{Force_Mono_EP}). This important
distinction is due to the existence of few much longer chains in the LPB.
Therefore, the reactive force of a LPB extends to larger separations and the
simulation predicts a non-vanishing force at $z > \langle z \rangle = 22.75$.

\begin{figure}[bht]
%\vspace{0.8cm}
\includegraphics[scale=0.30,angle=0]{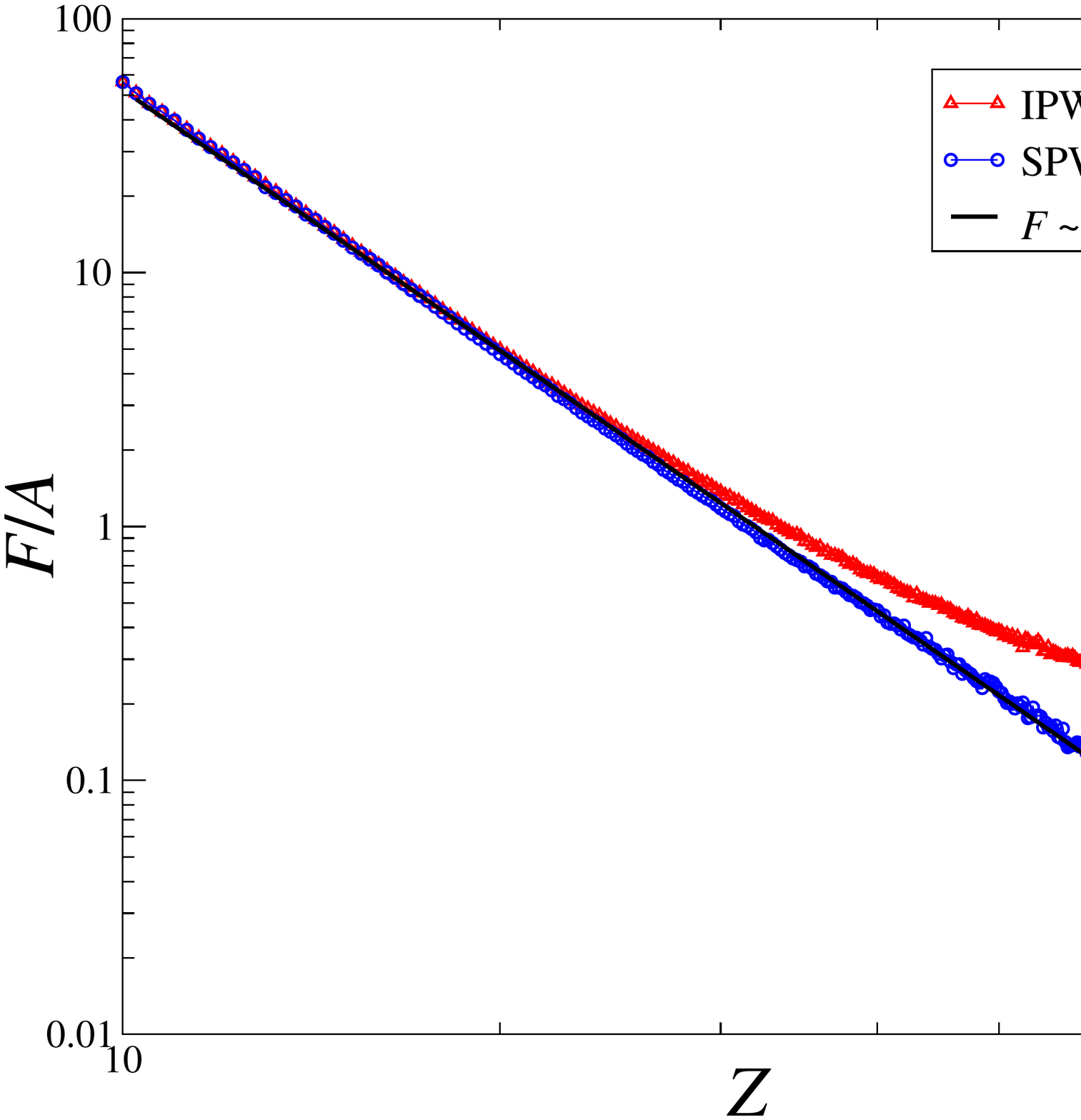}
\caption{Comparison between force-displacement curves of relaxed LPB,
obtained with both impermeable and semi-permeable wall for $\left \langle
L \right \rangle = 32$ at $\sigma_g = 1.0$.
\label{Force_Semi}}
\end{figure}

In order to be as close as possible to real applications of LPB, in which a LPB
is subjected to force measurement either by AFM or SFA instruments, we also
examined the case of a semi-permeable wall (SPW) as opposed to that of
completely impenetrable rigid wall (IPW). With a SPW,  a complete equilibration
of the whole LPB takes place under conditions when the unreacted single monomers
may move freely through the wall and enter the volume above it. In this way the
monomer concentration under the moving wall is not affected by the wall motion
and remains equal to that in the whole volume regardless of the wall position as
is the case with an AFM tip. Otherwise, any change in the wall position would
have induced a change in the average chain length $\langle L \rangle$ because of
growing $\phi_t$, according to Eq.~(\ref{eq:L_aver}) and, therefore, also in the
reactive force of the LPB even before the wall gets in touch with the brush.
With other words, any variation of an impenetrable wall would produce an
essentially different LPB which precludes a meaningful study of the force - wall
position relationship at fixed monomer concentration $\phi$. In order to avoid
inaccuracies and in view of the additional time the monomers need to move
through the SPW, we chose in this case  the equilibration periods twice longer
than in the IPW case with rate in the range of $10^{-8}-10^{-10}$.

The result of such simulation for force-displacement curve in loading experiment
is shown in Figure~\ref{Force_Semi}, together with the curve for the same LPB,
compressed by an IPW of identical area. The average chain length in both cases
is $\langle L \rangle = 32$ and the grafting density is $\sigma_g=1.0$. The
force exerted by the LPB on the SPW excludes the interaction between non-grafted
monomers and the test surface. One can immediately see that the SPW clearly
gives a weaker force compared to that with IPW at small compressions. This
difference becomes indistinguishable at lower separation or large degrees of
compression, and the curves in both cases superimpose perfectly for $z \leq 20$
as expected. It is evident from Figure~\ref{Force_Semi} that the
force-displacement curves appear as straight lines in logarithmic coordinates,
suggesting a power law scaling, $F \propto z^{-\alpha}$, where the observed
exponent $\alpha \approx 3.4$ is shown to  describe a dense LPB system both in
the SPW and IPW case.

\begin{figure}[htb]
\vspace{0.8cm}
\includegraphics[scale=0.4,angle=0]{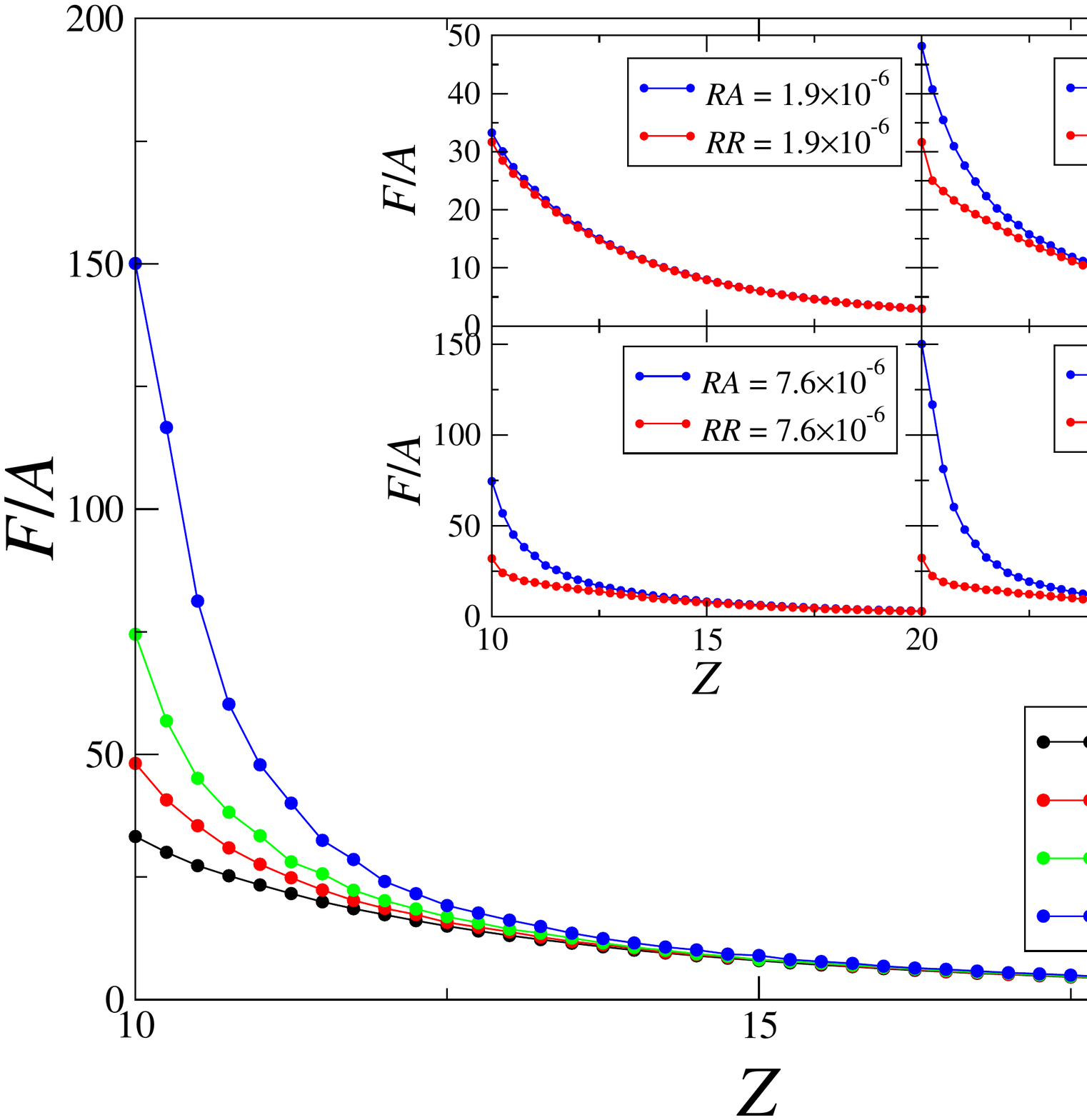}
\caption{The effect of moving velocity of the test wall on the force -
displacement curve of LPB with $\langle L \rangle = 32$ and $\sigma_g = 1.0$
during the loading process. In the inset a detailed view of the force hysteresis
during loading and unloading experiments for a LPB is shown. Results for four
incoming and retraction velocities are included, as indicated.
\label{Force_EP_Hyster}}
\end{figure}

Eventually, in Figure~\ref{Force_EP_Hyster} we present the effects of tip scan
rate on the hysteresis behavior in the force-displacement curve during a
loading/unloading experiment. Recently, the response of a compressed
monodisperse polymer brush as function of $\sigma_g$ and compression rate
$RA/RR$ was studied by Brownian Dynamics simulation \cite{Elvingson}. Our
simulations for a LPB system are performed
with an average chain length $\langle L \rangle=32$ and grafting density
$\sigma_g=1.0$. The main panel of Figure~\ref{Force_EP_Hyster} shows the
observed force-displacement curve in loading experiments for different
indentation/compression rate. Evidently, as the scan rate gets faster, the final
values of the response force at small separations $z<15$ grow while the
superposition region is shifted to lower compressions. These predictions are
found to agree well with experimental interpretations of Hoh and Engel
\cite{Hoh} in their AFM indentation test even though our approaching rates
are about three order of magnitude smaller than in a real AFM indentation
experiment. In the inset plots of Figure~\ref{Force_EP_Hyster} one may see
the hysteresis becomes stronger with increasing loading / unloading rate which
reflects also growing energy dissipation given by the enclosed area of
the indentation and retraction curves. The values of energy dissipation has
been measured experimentally for different systems \cite{Treventhan}. As can be
seen from the Figure~\ref{Force_EP_Hyster}, by decreasing the scan rate the
hysteresis may be completely eliminated when full equilibrium is reached without
energy dissipation. The latter shows an elastic behavior of a thin layer in
which the material can regains step by step its own shape
during the withdrawal process \cite{Lee}.

In practice, the sample may loose elasticity in the process of compression so
that when the tip is withdrawn, it does not regain immediately its former
shape. The load gradually decreases whereas the penetration depth stays the
same. Hence indentation and retraction curves seldom overlap. At a given
penetration depth the force of the unloading curve is smaller than the force at
loading. This difference between the indentation and the withdrawal force curves
appears as a ``loading-unloading hysteresis''. The hysteresis may lead to an
incorrect determination of displacements and, in particular, because of the
hysteresis, the load in the loading curve for a given displacement may appear
bigger and finally overcome those of the unloading curve. Experimentally, Hoh
and Engel \cite{Hoh} have shown that the loading/unloading hysteresis is
scan~rate- (or, velocity)-dependent. At high scan rates the separation between
the contact lines in indentation and retraction modes is large and a
considerable hysteresis appears. As the scan rate is decreased, the system has
more time to recover and this separation reaches a minimum so that the
hysteresis becomes smaller and may vanish.

\section{Comparison Between Experiment and Simulation}
\label{sec_comparison}

Here we focus on the comparison of our simulations exclusively with
own AFM-based, colloidal probe compression measurements. PNIPAAm brush was
grafted on the surface of a self-assembled monolayer containing the
initiator, using surface-initiated atom transfer radical polymerization (ATRP).
Varying the reaction time and monolayer initiator concentration controlled the
molecular weight and grafting density, respectively. For details on the
synthesis of PNIPAAm brush samples and evaluation of grafting density see
Supporting information. AFM measurements were performed on ATRP grown PNIPAAm
films in a standard fluid cell containing deionized water as a good solvent at
$26~{\rm C^\circ}$ which was below the LCST temperature of PNIPAAm.

Experimentally, a surface force apparatus (SFA) was used by Zhu {\it et.
al.},\cite{Zhu} and Plunket {\it et. al.},\cite{Plunket} to measure the force
encountered between tethered poly({\it N}-isopropylacrylamide) (PNIPAAm) at,
below, and above lower critical solution temperature (LCST) and plain mica
surfaces and mica surfaces coated with lipid bilayers. In the SFA technique, the
force of interaction between two perpendicular crossed cylinders can be measured
as a function of the distance between the two cylindrical surfaces. Plunkett
{\it et. al.},\cite{Plunket} coated the surface of one cylinder with terminally
anchored PNIPAAm and kept the surface of the other cylinder bare.

In our simulation, force was measured between a LPB and a wall, taken either
as a hard impenetrable, or a semi-permeable layer, and compared to experimental
data obtained for PNIPAAm under good solvent condition (below LCST) as well as
to
the simulation results for monodisperse MPB case. This comparison makes sense
only when the length of chains in the MPB brush, $N$, is equal to the mean chain
length $\langle L \rangle$ of the LPB. However, for a living polymer $\langle L
\rangle$ depends essentially on the concentration of free monomers in the
container \cite{Wittmer2}, and in the case of impenetrable wall will therefore
change dynamically with the variation of monomer concentration as the wall is
moved with respect to the grafting plane. To prevent this and examine solely
the effect of polydispersity at equal length of the chains in both LPB and MPB
brush, we use in our simulation also a semi-permeable wall that is impenetrable
for the grafted polymers yet completely permeable for the single non-polymerized
monomers. These monomers can then move freely through the mobile wall and thus
keep the total concentration in the box constant. \begin{figure}[ht]
\includegraphics[width=8cm, height=6cm]{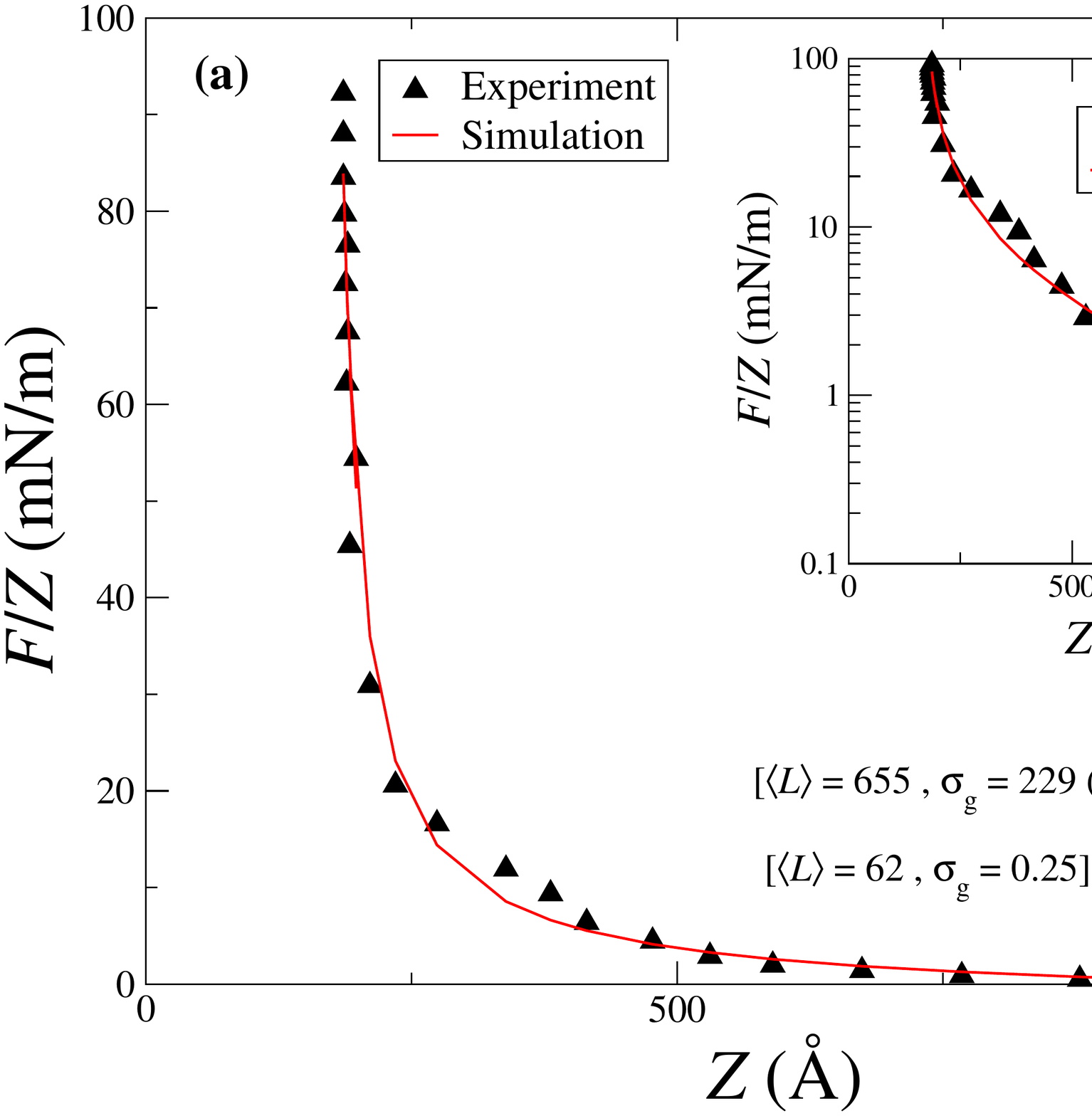}
\hspace{1cm}
\includegraphics[width=8cm, height=6cm]{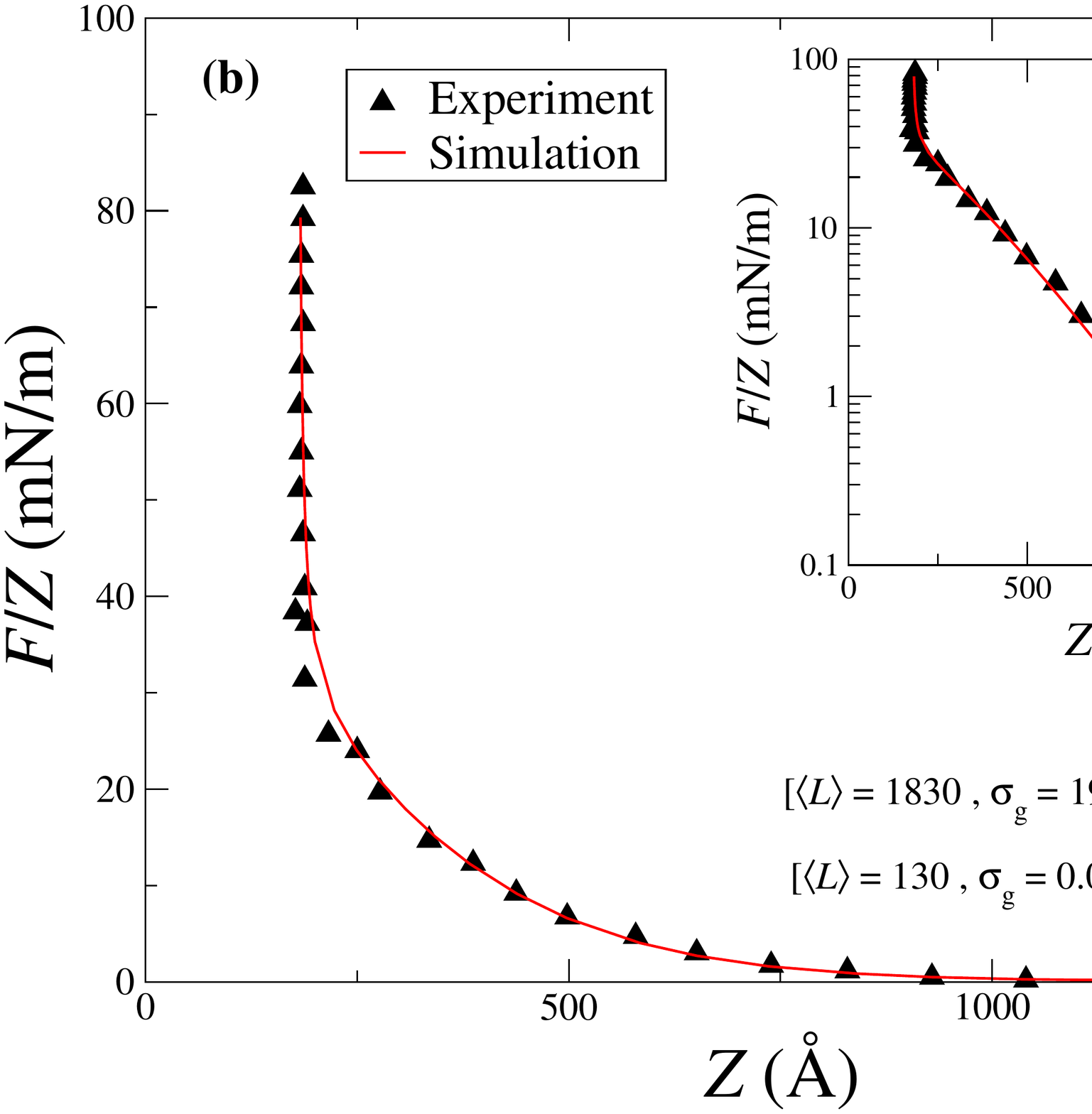}
\caption{(a) Normalized force of Plunkett {\it et. al.},\cite{Plunket} for
$\langle L\rangle^{\rm exp} =$ 655 plotted vs the distance between
untreated mica and a PNIPAM brush polymerized at grafting density $229$
\AA{}$^2$/chain and comparing to simulated interaction force for a given
grafting density and degree of polymerization, as indicated. (b) Same as (a) but
for $\langle L \rangle^ {\rm exp} = 1830$, grafting density $1930$
\AA{}$^2$/chain and simulated interaction force for a different set of grafting
density and degree of polymerization, as indicated. \label{Exper-Simul}}
\end{figure}

Figure~\ref{Exper-Simul} illustrates the starting point of our analysis: from
the data presented by Plunkett {\it et. al.}, \cite{Plunket} for a range of end
grafted PNIPAM polymers that were synthesized by living polymerization (ATRP) on
a gold flat surface with different grafting density (${\sigma_g}^{\rm exp}$) and
degree of polymerization ($\left<L\right>^{\rm exp}$), we can identify a number
of choices (${\sigma_g}^{\rm exp}$, $\left<L\right>^{\rm exp}$) where a
quantitatively precise mapping of the experimental force-displacement curve on
the simulated interaction force, for carefully chosen pairs of parameters
(${\sigma_g}^{\rm sim}, \left<L\right>^{\rm sim}$), is possible. Here we use
subscripts ``exp'' and ``sim'' to distinguish the real experimental data from
their simulation counterparts, respectively. Thus, apart from suitable
adjustment of pair parameters, the length scale of the simulation (i.e., the
lattice spacing) $s$ is adjusted to physical units by requiring that
\begin{eqnarray}\label{eq:fit}
\left(\frac{s}{2\left<R_F^2\right>^{1/2}}\right)^{\rm exp} =
\left(\frac{s}{2\left<R_F^2\right>^{1/2}}\right)^{\rm sim}
\end{eqnarray}
The results of this fitting process for two experimentally calculated values
$\left({s}/{2\left<R_F^2\right>^{1/2}}\right) = 0.033$ and $0.046$ (obtained by
Plunkett {\it et. al.} \cite{Plunket}) are shown in Figure~\ref{Exper-Simul}. As
can be seen from the figure, a good fit over the entire range of separations $z$
is obtained for $\left<L\right> = 62$ and $\sigma_g = 0.25$ in case (a) and for
$\left<L\right> = 130$ and $\sigma_g = 0.0625$ in case (b). The corresponding
experimental values for $\left<L\right>$ and $\sigma_g$ are indicated in the
figure. It should be noted, from experimental point of view, that all
experimentally determined grafting densities are sufficiently high
\cite{Plunket} so that the excluded volume interactions make the chains extend
away from the substrate and form a brush structure. Therefore, the selected
experimental grafting density for the comparison with simulated data at $\phi =
0.5$ (which corresponds to semi-dilute brush regime in the simulation) confirmed
that the experimentally synthesized grafted layer is in the genuine brush
regime.

\begin{figure}[bht]
\includegraphics[width=8cm,
height=6cm]{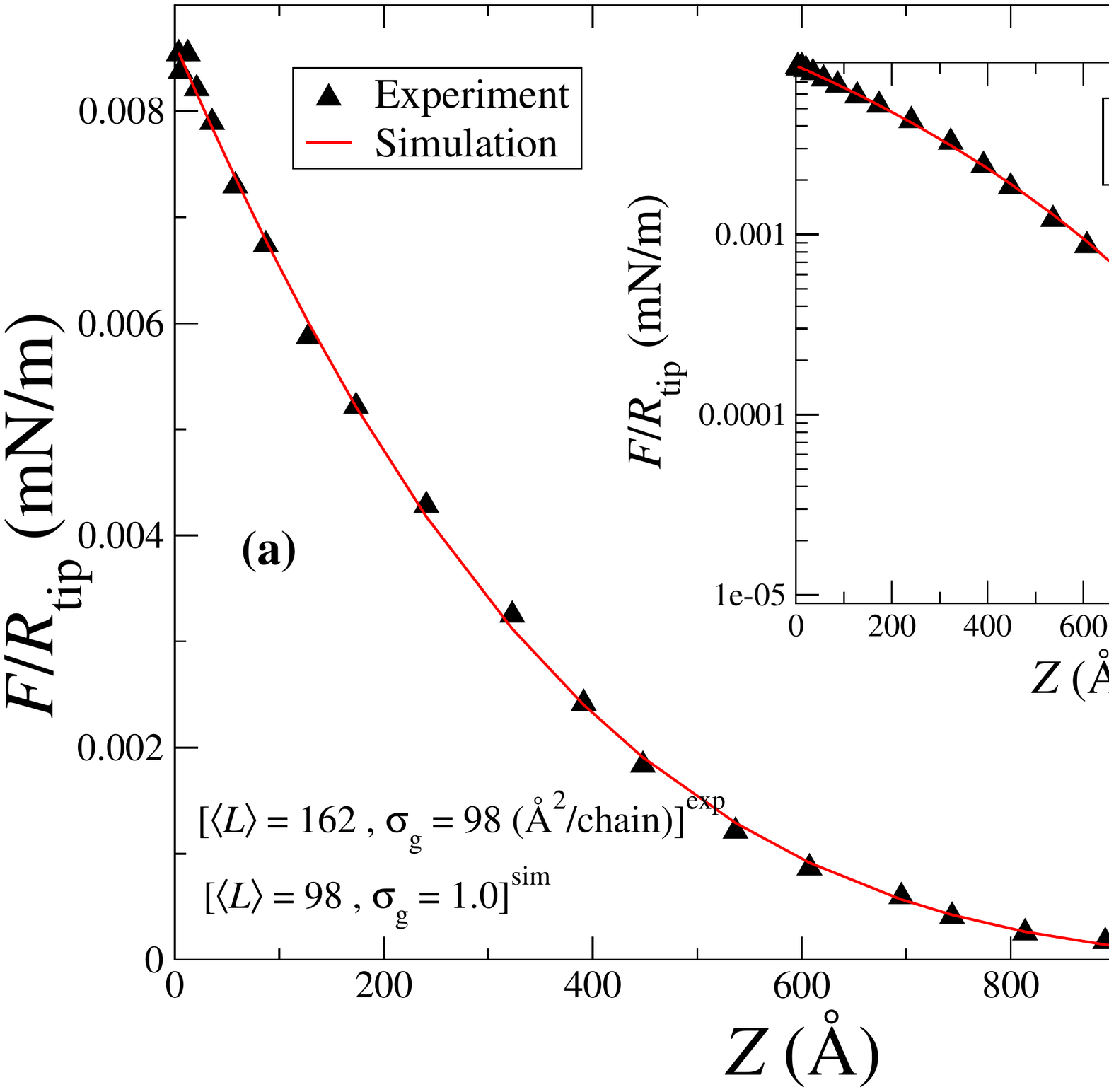}
\hspace{1cm}
\includegraphics[width=8cm,
height=6cm]{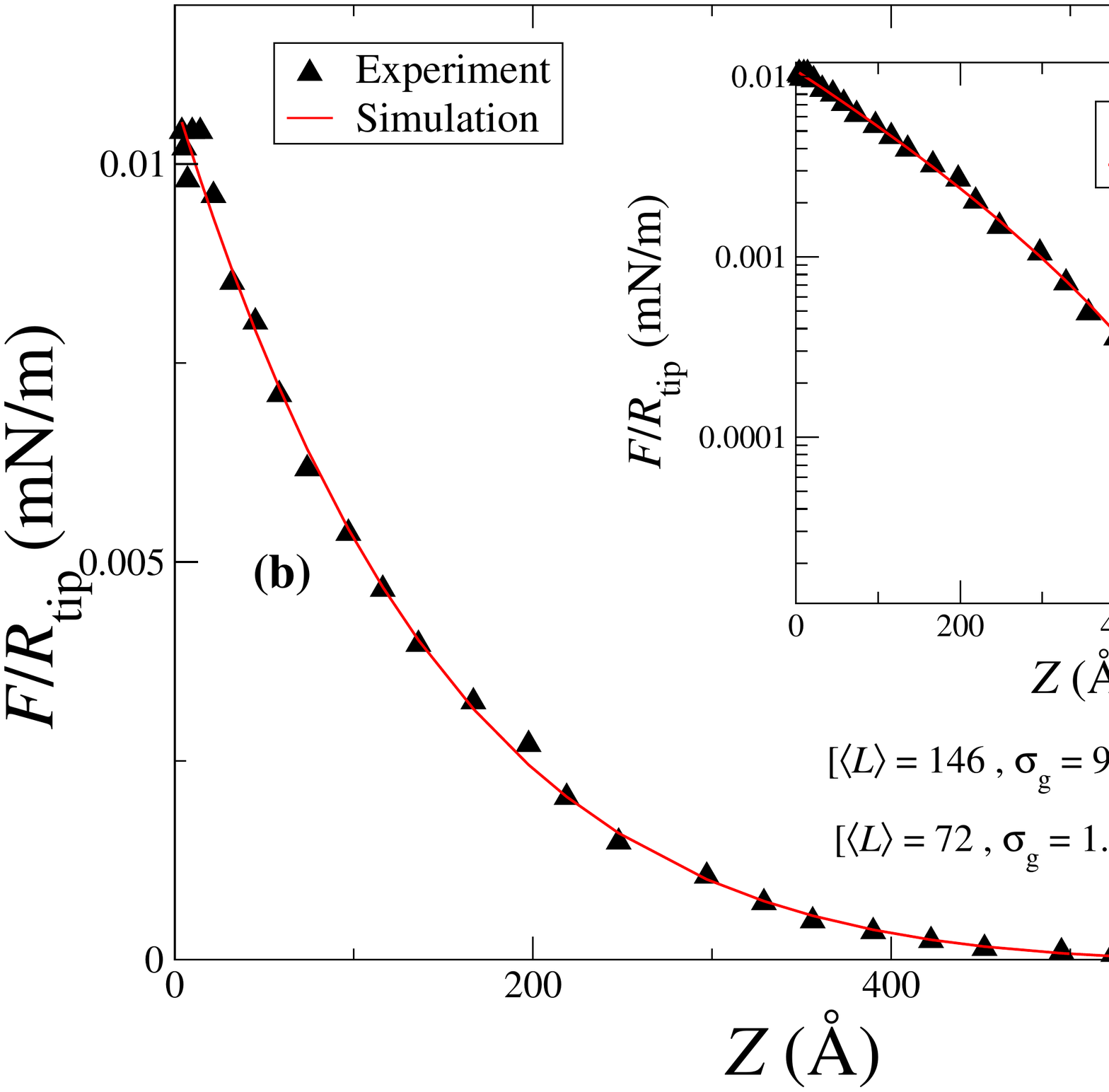}
\caption{(a) Normalized force for $\langle \it
{L}\rangle^{\rm exp} =$ 162 plotted vs the distance between AFM colloidal tip
and a PNIPAM brush polymerized at grafting density $98$ \AA{}$^2$/chain
and comparing to simulated interaction force for a given grafting density and
degree of polymerization, as indicated. (b) Same as (a) but for $\langle
L \rangle^ {\rm exp} = 146$, grafting density $98$ \AA{}$^2$/chain and
simulated interaction force for a different set of grafting density and
degree of polymerization, as indicated.
\label{Exper-Simul-OwnData}}
\end{figure}

Bearing in the mind that the self-similar character of the conformation of the
chains allows for some arbitrariness of the scales, one has to do an explicit
mapping of a coarse-grained model of a flexible linear polymer either in grafted
or non-grafted form on the real material. The principles of this mapping
originate from the assumed general shape of the free energy of a chain
$f_p$, when one compares it with the properties of ideal chains in
good solvents with an end-to-end distance $R$. The free energy of a single
chain for a regular solution is obtained by inserting the excluded volume
contribution $f_p^e$ and entropic component $f_p^s$ into the equation $f_p =
f_p^e + f_p^s$. The excluded volume contribution to the free energy of a single
chain is obtained by integration over the occupied volume, hereby
calculating the average over the all conformations
\begin{eqnarray}
f_p^e = \frac{1}{2}\int kT\upsilon_e\left<c_m^2\left(\bf r\right)\right>d^3\bf r
\end{eqnarray}
where $\left<c_m^2(\bf r)\right>$ is square average local monomer density and
$\upsilon_e$ is excluded volume interaction parameter. If we choose for the
description of the mean local density $\left<c_m\left(\bf r\right)\right>$ a
Gaussian function, with a radius of gyration $R_g$, we obtain
\begin{eqnarray}
f_p^e = \frac{kT}{2}\upsilon_eN^2\left(\frac{3}{4\pi R_g^2}\right)^{3/2}
\end{eqnarray}
Eventually, one may choose the ideal state with vanishing excluded volume
forces and coil size in thermal equilibrium $R_0$, where $R_g=R_{g,0}$ and
$R_0^2=a_0^2N$, and replace $R_g$ by $R$ assuming $R \sim R_g$. Thus one may
express $f_p^e$ as
\begin{eqnarray}\label{eq:f_p_e}
f_p^e = \frac{kT}{2}3^{3/2} \psi\left(\frac{R_0}{R}\right)^3
\end{eqnarray}
The parameter $\psi=\left(3/2\pi\right)^{3/2}\left(\upsilon_e/a_0^4\right)R_0$,
is dimensionless and determines the excluded volume energy associated with a
single chain.

A further requirement is that one needs an expression to take into account the
conformational entropy indicating the elastic force built up on a coil
expansion. This equation yields the second part of the free energy, $f_p^s$, and
is given by \begin{eqnarray} \label{eq:f_p_s} f_p^s\approx
kT\left(\frac{R}{R_0}\right)^2 \end{eqnarray} Finally, combination of the
Eqs.~(\ref{eq:f_p_e}) and (\ref{eq:f_p_s}) yields the free energy of a chain as
a function of $R$. One may determine the important relation between $R_F$ and
the degree of polymerization $N$ by calculating the equilibrium value of $R$ at
the minimum of the free energy $f_p$, where $df_p/d(R/R_0) = 0$. $R_0$ is the
coil size in thermal equilibrium. This leads to the relation $R \simeq \left (
\nu_e / a_0^4 \right )^{1/5} R_0^{6/5} = \left ( \nu_ea_0^2 \right )^{1/5}
N^{3/5}$. One may identify $R$ with the Flory radius $R_F$ and it can be
represented thus as $R_F = a_F N^{3/5}$ with $a_F \backsimeq \left ( \nu_ea_0^2
\right)^{1/5}$.

The scaling law Eq.~(\ref{eq:fit}) is indeed in full accord with experiments
(e.g., see ref 94) and it can be used  for the
translation of experimental findings to results of simulation in the case of
various expanded polymer system in a good solvent. Of particular interest, we
applied the same criteria in order to map our coarse-grained level simulations
of force-displacement curve of a LPB to the experimental results  obtained by
SFA experiment (see Figure~\ref{Exper-Simul}) and own colloidal probe AFM
analysis on PNIPAAm brush in the good solvent. Since in the case of LPB system,
there is only a single factor relating the length scale of the simulation to the
real length scale and leads to a mapping process that conserves the chain
conformation during translation, it can be expected that the precise description
of coarse-grained model for force-displacement curve in Figure~\ref{Exper-Simul}
is restricted to special cases. As an example, Figure~\ref{Exper-Simul-OwnData}
presents the force-displacement behavior in a AFM analysis for the pairs
($\left<L\right>=162,~\sigma_g=98$ \AA$^2$/chain)$^{\rm exp}$, and
($\left<L\right>=146,~\sigma_g=98$ \AA$^2$/chain)$^{\rm exp}$. All force data
are normalized by the colloidal tip radius $R_{\rm tip}$. As can be seen from
Figure~\ref{Exper-Simul-OwnData}, we find the closest corresponding translated
pairs in simulation data as ($\left<L\right>=98,~\sigma_g=1.0$)$^{\rm sim}$, and
($\left<L\right>=72,~\sigma_g=1.0$)$^{\rm sim}$. The inset in
Figure~\ref{Exper-Simul-OwnData}(a) shows that for very small compressions, the
semi-log plot of the force-displacement curve deviates somewhat from the
coarse-grained model and near $z \approx 900$ {\AA} the simulation curve
$\left<L\right>^{\rm sim}=98$ slightly underestimates the experimental data. As
shown in the inset to Figure~\ref{Exper-Simul-OwnData}(b), in the case of lower
degree of polymerization, $\left<L\right>^{\rm sim}=72$, the same deviation is
also observed for lower separations while the underestimation is slightly
decreased.

\section{CONCLUSIONS}\label{sec_summary}

In the present work we have studied and tried to reproduce the force exerted by
a LPB chains on a test surface. Our studies have been carried out by means of an
efficient off-lattice Monte Carlo algorithm. To this end comprehensive Monte
Carlo simulations of the properties of a LPB made of flexible living polymer
chains under good solvent conditions have been carried out. Most of our studies
consider moderate and high grafting densities of the chains whereby the ensuing
polymer brush may be classified as strongly stretched.

An examination of the static properties of such LPB suggests that the observed
monomer density profiles reveal power-law decline of the density $\phi(z)$ with
increasing distance $z$ perpendicular to the grafting plane, $\phi(z) \propto
z^\alpha$ with $\alpha \approx 0.64$ which is very close to the theoretically
predicted value of $\alpha = 2/3$ for diffusion-limited aggregation
\cite{Wittmer1}. The probability distribution $c(N)$ of chain lengths $N$ in the
strongly polydisperse living polymer brush is also found to follow a power law
relationship, $c(N) \propto N^{-\tau}$, which qualitatively differs from the
exponential Flory-Schulz Molecular Weight Distribution typical for living
polymers in the bulk. The observed value of $\tau \approx 1.70$ is again rather
close to the value of $\tau = 7/4$, predicted for the case of ``needle growth``
(Diffusion-Limited Aggregation Without Branching) \cite{Wittmer1}, except for
cases of rather short LPBs, grown at high temperature $T > 1.0$, where the
chains do not stretch strongly and an exponential MWD is observed. One may thus
conclude that the static properties of a LPB at equilibrium are very different
from those of a semi-dilute or dense solution of living polymers in the bulk,
and are governed by different laws.
 
One of the main concerns of this work has been the study of the force, exerted
by a LPB on a test surface, and the comparison of simulational data with that
from our own and other laboratory experiments. We have compared the
force-displacement behavior from the set of experiments \cite{Plunket}, obtained
by SFA analysis on PNIPAAm brush under good solvent condition at temperature
below LCST of PNIPAAm (Figure~\ref{Exper-Simul}), and also our own experimental
results obtained by colloidal probe AFM measurements on PNIPAAm brush at same
conditions of SFA (Figure~\ref{Exper-Simul-OwnData}) with data from the present
simulation. Generally, by adjusting the conformational parameter $s/2\langle
R_F^2\rangle^{1/2}$ and the conversion factor from model lattice spacings to
realistic nanometer scale, we find an almost perfect agreement. Small deviations
between simulation results and experiment are only found for weak compression
where in the case of SFA measurements, the simulation slightly overestimates the
data and in the case of AFM experiments, the simulation slightly underestimates
them.

It is clear, however, that more work is needed until full understanding of the
properties of living polymer brushes, obtained by ATRP is achieved. Especially
interesting and unexplored is the problem of LPBs, grown on curved surfaces,
including the corresponding force-distance relationship which we see as our
next target for research.

\section*{Acknowledgments}

K. Jalili and A. Milchev are indebted to the Max-Planck Institute for Polymer
Research in Mainz, Germany, for hospitality and support during their work on
the problem. A. M. gratefully acknowledges financial support by CECAM nano~SMSM.

\section*{Supporting Information Available}

Full details of the experimental characterization used in this study are
provided as a file in Supporting Information. This material is available free of
charge via the Internet at http://pubs.acs.org.

%\newpage

\newpage

\begin{figure}[bht]
\vspace{1.5cm}
\includegraphics[scale=0.5,angle=0]{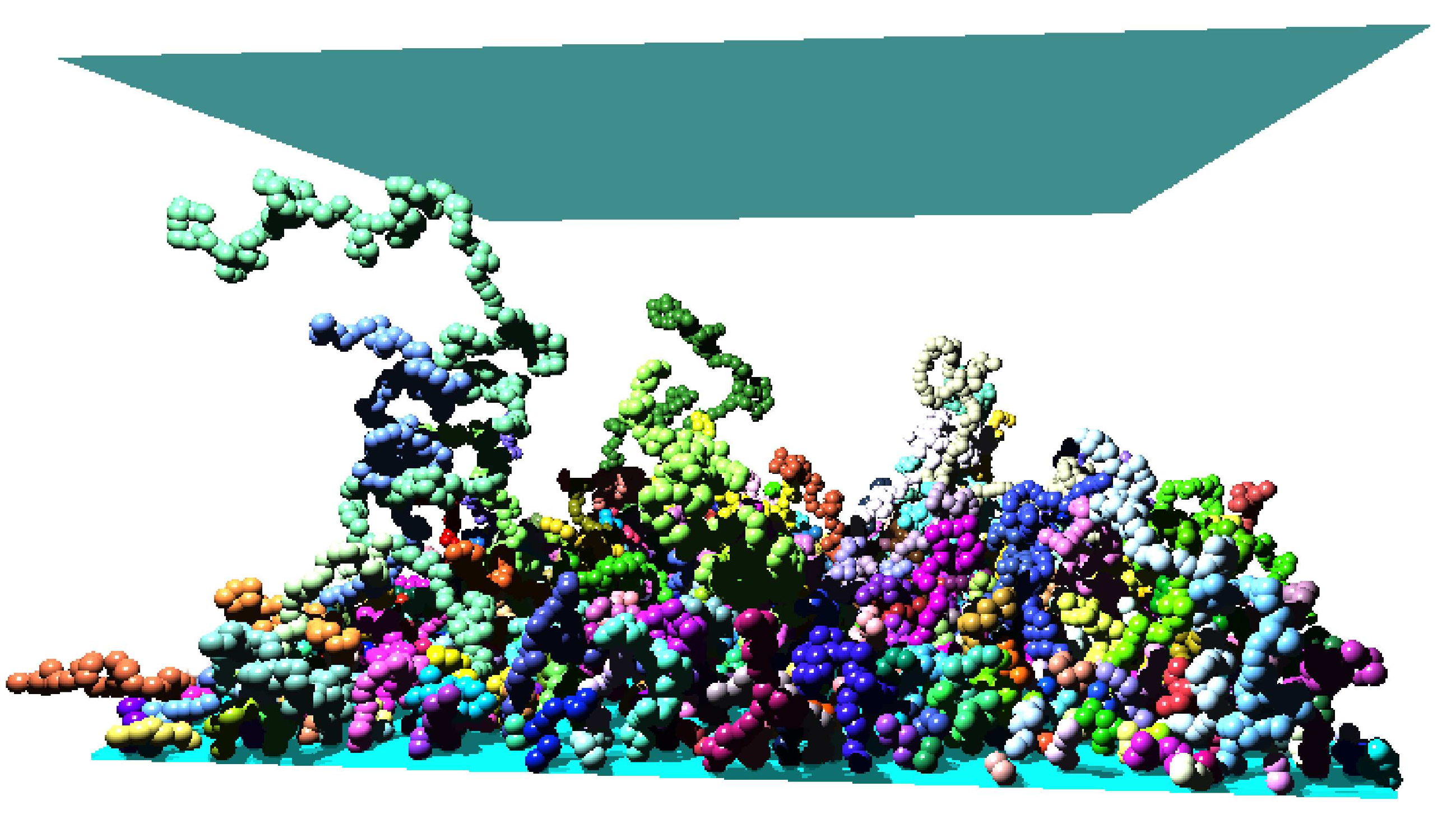}
%\caption{

Table of Contents Graphic:\\
''Dynamic Compression of {\it in situ} Grown Living Polymer Brush: Simulation
and Experiment'', \\
by K. Jalili, F. Abbasi, and A. Milchev
%}
\author{K. Jalili$^{1,2}$, F. Abbasi$^2$, A. Milchev}

\end{figure}

\end{document}